\documentclass{elsart}

 \usepackage{graphicx}

\usepackage{amssymb}

\begin{document}

\begin{frontmatter}



\title{Neutrino Masses and Oscillations:\\
 Triumphs and Challenges}

\author{R.~D.~McKeown and P.~Vogel}

\address{W.~K.~Kellogg Radiation Laboratory, California Institute of
Technology, Pasadena, California 91125, USA}


\begin{abstract}
The recent progress in establishing the existence of finite neutrino masses and mixing between 
generations of neutrinos has been remarkable, if not astounding. 
The combined results from studies of atmospheric neutrinos, 
solar neutrinos, and reactor antineutrinos paint an intriguing picture for theorists and 
provide clear motivation for future experimental studies. 
In this review, we summarize the status of
experimental and theoretical work in this field and explore the future opportunities 
that emerge in light of recent discoveries.

\end{abstract}

\end{frontmatter}

\section{Introduction and Historical Perpective}
For almost 70 years, neutrinos have played a pivotal role in the
the quest to understand elementary particles and their interactions.
The neutrino was actually the first particle proposed by a theorist
to guarantee the validity of a symmetry principle:
Pauli boldly suggested \cite{Pauli} that neutrinos 
(invisible then and remaining so for another twenty five
years) are emitted together with electrons in nuclear
beta decay to salvage both energy and angular momentum conservation
in the beta decay process.
Experimental confirmation of Pauli's hypothesis required many 
decades of experimental effort,
but the discovery of the antineutrino in the fifties 
by Reines and Cowan \cite{Reines}
and subsequent experiments in the early sixties
ultimately led to the Nobel Prizes for Reines (1995) and
Lederman, Schwartz and Steinberger (1988). More recently, the 2002 Nobel prize
was awarded to Davis and Koshiba for their seminal roles in the development
of neutrino astrophysics (along with Giaconni for x-ray astrophysics). 

The standard model of electroweak interactions,
developed in the late 1960's, provided a theoretical
framework to incorporate the neutrinos as left-handed partners to the charged leptons,
organized in two generations along with the quarks.  
The subsequent discovery of charmed quarks
and the third generation of quarks and leptons completed the modern view of the standard 
model of electroweak interactions. 
This version possessed additional richness to incorporate
CP violation, and further efforts to unify the strong interaction led to the development
of Grand Unified Theories (GUTs). These GUTs provided a natural framework for nucleon
decay and for neutrino masses, and motivated many experiments in the field.

Following on the successes of big-bang nucleosynthesis (BBN) 
and the discovery of the cosmic
microwave background, it also became clear that neutrinos 
were potentially major players in the history of the
early universe. The increasing evidence for the existence 
(and then dominance) of dark matter
in the universe then led to the economical and seductive 
hypothesis that neutrinos, with small but finite
mass, could provide the mass to explain the dark matter. This set the stage for a
major experimental assault on the issue of neutrino mass and its role in cosmology, 
and provided substantial impetus to
a worldwide program of experiments addressing the issues of finite neutrino mass and
the possibility of mixing between generations.

Although it now appears that neutrinos are not likely 
the source of dark matter in the universe, the
experimental evidence obtained in the last decade
for finite neutrino masses and mixing between generations is strong
and irrefutable. The pattern of masses and mixing angles emerging 
from the experiments provides
an intriguing glimpse into the fundamental source of particle mass 
and the role of flavor 
in the scheme of particles and their interactions. 
The scale of neutrino mass differences motivates new 
experimental searches for double beta decay 
and end-point anomalies in beta decay, as well as
new studies of oscillation phenomena using accelerators, nuclear reactors, and 
astrophysical sources of neutrinos.

In this review, we attempt to synthesize these themes, 
present a concise and coherent view
of the experimental and theoretical developments leading to the current picture, and
motivate the future explorations necessary to resolve the remaining issues
\footnote{The reference list is, for reasons of brevity, far from complete.
We apologize to those whose work is not quoted, and refer to numerous reviews
and monographs to provide a more complete reference list.}.
We begin with a summary of the theoretical motivation 
for studying neutrino masses and mixing, 
along with a development of the phenomenological 
framework for interpreting the experiments. 
This includes
neutrino oscillations in vacuum as well as in matter, CP violation, ordinary beta decay, and double
beta decay. We then review the recent experimental data that contribute to our present knowledge of
these neutrino properties. We discuss the successful neutrino
oscillation measurements, including the key contributions from various solar neutrino
experiments, cosmic ray-induced atmospheric neutrino studies, and the recent dramatic
results from the Sudbury Neutrino Observatory and the KamLAND reactor neutrino experiment.
Important additional information is obtained 
from the experimental attempts
that have thus far only yielded limits on neutrino masses
such as double beta decay and tritium beta
decay. We also will briefly discuss the role of massive 
neutrinos in cosmology, and the corresponding constraints.

The discussion of the future experimental program is separated into two different time
scales.
In the near-term future, the planned experiments will focus on a
better determination of the mixing matrix
parameters, in particular $\theta_{13}$,
and resolution of the LSND puzzle (right or wrong). Further studies in the longer-term 
will of course depend on the outcome of these measurements, but there
is substantial interest in pursuing the search for CP violation, the possibility
that neutrinos are Majorana particles and other issues related to the mechanism(s) responsible
for the observed phenomena.

\section{Theoretical Framework}

It is natural to expect that neutrinos have nonvanishing masses, since
they are not required to be massless by gauge invariance or other
symmetry principles. Moreover, all other known fermions, quarks and
charged leptons, are massive. Nevertheless, neutrinos are very light,
much lighter than the other fermions, and this striking qualitative
feature needs to be understood, even though we do not know why e.g. the
electron mass is what it is known to be, and why muons and taus are
heavier than electrons.

The most popular explanation of the small neutrino masses is the 
``see-saw mechanism'' \cite{seesaw} in which the neutrino
masses are inversely proportional to some large mass scale
$M_{\mathcal H}$. Simply put, if neutrinos are Majorana particles,
i.e., indistiguishable from their own antiparticles, they have
only two states, corresponding to the two possible spin orientations.
On the other hand, Dirac particles, distinct from their antiparticles,
have four states, two spin orientations for the particle and 
antiparticle. In the  ``see-saw mechanism'' there are still
four states for each neutrino family, but they are
widely split by
the ``Majorana mass term'' (explained later) into the 
two-component light neutrinos with masses $M_{\mathcal L}$ and the
very heavy (sterile) two-component neutrinos with 
the mass $M_{\mathcal H}$,
in such a way that $M_{\mathcal L}M_{\mathcal H} \cong M_{\mathcal D}^2$.
Further, if we assume that the ``Dirac mass'' $ M_{\mathcal D}$
is of the same order of magnitude as the Dirac fermion masses 
(masses of quarks and charged leptons), we can understand why
$M_{\mathcal L}$ is so small, provided $M_{\mathcal H}$ is very large.
If that explanation of the small neutrino mass is true, then 
the experimentally  observed neutrinos are Majorana particles,
and hence the total lepton number is not conserved. Observation
of the violation of the total lepton number conservation (we explain
further why it is difficult to observe it) would be a signal
that neutrinos are indeed Majorana particles.

Neutrinos interact with other particles only by weak interactions
(at least as far as we know); they do not
have electromagnetic or color charges. 
That further distinguishes them from the
other fermions which also interact electromagnetically
(charged leptons and quarks) and strongly (quarks).
  
It is well known that the weak charged currents of quarks do not
couple to a definite flavor state (or mass state), but to
linear combinations of quark states. This phenomenon is described
by the Cabibbo-Kobayashi-Maskawa (CKM) matrix, which by convention
describes linear combinations of the charge $-e/3$ quark mass
eigenstates $d,s,b$. The mixed states $d',s',b'$, form weak
doublets with the $u,c,t$ quark mass eigenstates.
Consequently, the three generations of quark doublets 
are not independently preserved in charged-current weak
processes.

An analogous situation is encountered with neutrinos. There, however,
one can directly observe only the `weak eigenstates', i.e., neutrinos
forming weak charged currents with electrons, muons or tau 
\footnote{The analogy is not perfect. For quarks, the
`mass eigenstates' are also `flavor eigenstates', with labels
$d,s,b$. Since the CKM matrix is nearly diagonal, the labels
$d',s',b'$ can be used for the `weak eigenstates'.
In neutrinos, on the other hand, only the `weak eigenstates'
have special names, $\nu_e,\nu_{\mu},\nu_{\tau}$ while the mass
eigestates are labelled $\nu_1,\nu_2,\nu_3$ and cannot be 
associated with any particular lepton flavor.}.
Information 
on the  neutrino mixing matrix is best obtained by the study of
neutrino oscillation, a quantum mechanical interference effect resulting from
the mixing. As a consequence, like in quarks, the flavor (or family)
is not conserved and, for example, a beam of electron neutrinos could, after
travelling a certain distance, acquire a component of muon or tau
flavor neutrinos. We describe the physics and formalism of the
oscillations next.   The formalism has been covered in great detail
in numerous books and reviews. Thus, we restrict ourselves
only to the most essential points.
The interested reader can find more details in the monographs 
\cite{BV,Boris,Mohapatra,Caldwell,Bahcall,Winter,Raffelt}
and recent reviews \cite{Fisher,EV,BMW,G-GN,BGV,Altarelli,BGGM,King}.

\subsection{Neutrinos in the standard electroweak model}

In the standard model individual lepton charges ($L_e = 1$ for
$e^-$ and $\nu_e$ and $L_e = -1$ for
$e^+$, and $\bar{\nu}_e$ and analogously for $L_{\mu}$ and $L_{\tau}$)
are conserved. Thus, processes such as $\mu^+ \rightarrow e^+ + \gamma$,
or $K_L \rightarrow e^{\pm} + \mu^{\mp}$ are forbidden. Indeed, such
processes have not been observed so far, and small upper limits
for their branching ratios have been established.

Based on these empirical facts, the standard model places the 
left-handed components of the charged lepton and neutrino fields
into the doublets of the group $SU(2)_L$,
\begin{equation}
\psi_{\ell L} = \left( \begin{array}{c}
\nu_{\ell L} \\ \ell_L \end{array} \right) ~,~
\ell = e, \mu, \tau ~,
\label{e:double}
\end{equation}  
while the right-handed components of the charged lepton fields are
singlets. The right-handed components of the neutrino fields are
{\it absent} in the standard electroweak model by definition. 

As a consequence of this assignment, neutrinos are deemed to be massless, and the
individual lepton numbers, as well as the total one, are strictly conserved.
Note that with these assignments neutrino masses do not arise
even from loop corrections.
Thus, observation of neutrino oscillations, leading to
the neutrino flavor nonconservation, signals deviations
from this simple picture, and to the `physics beyond the standard model'. 

Note also that studies of $e^+e^-$ annihilation at the $Z$-resonance
peak have determined the invisible width of the $Z$ boson, caused by
its decay into unobservable channels. Interpreting this width
as a measure of the number of neutrino active flavors, one obtains
$N_{\nu} = 2.984 \pm 0.008$ from the four LEP 
experiments \cite{pdg}. We can, therefore, quite confidently
conclude that there are just three active neutrinos with masses of
less than $M_Z/2$. (The relation of this finding to the fact that
there are also three flavors of quarks is suggestive, but so far not
really understood.) Besides these three active neutrino flavors
there could be other neutrinos which do not partipate in weak interactions.
Such neutrinos are called `sterile'. In general, the active and sterile
neutrinos can mix, and thus the sterile neutrino can interact,
albeit with a reduced strength.  

Big Bang Nucleosynthesis (BBN) is sensitive to the number of neutrino 
flavors which are ultrarelativistic at the `freezeout' when the 
$n \leftrightarrow p$ reactions are no longer in equilibrium
and therefore it is perhaps sensitive to (almost) sterile neutrinos.
However, analysis of BBN (i.e., of the abundances of $^4$He, $d$, $^3$He, and
$^7$Li) are consistent with 3 neutrinos, and  disfavors 
$N_{\nu}=4$ for fully thermalized neutrinos \cite{Barger}.

\subsection{Neutrino mass terms}

In a field theory of neutrinos the mass is determined by the mass
term in the Langrangian. Since the right-handed 
neutrinos are absent in the standard electroweak model,
one can either generalize the model and define the mass term by using the
ideas of the see-saw mechanism, or one can add more possibilities
by adding to the three known neutrino fields $\nu_{\ell L}$
new fields, corresponding to possibly heavy right-handed 
neutrinos $\nu_{jR}$. The mass term is then
constructed out of the fields  $\nu_{\ell L}$, their charge-conjugated
and thus right-handed fields $(\nu_{\ell L})^c$ and from the fields
$\nu_{jR}, (\nu_{jR})^c$.
There are, in general, two types of mass terms,
\begin{equation}
-L_M = {\mathcal M}_{i \ell}^D \bar{\nu}_{iR} \nu_{\ell L}  + 
\frac{1}{2}{\mathcal M}_{ij}^M (\bar{\nu}_{iR})(\nu_{jR})^c + {\rm H.c.} ~.
\label{e:mass}
\end{equation}
(The term with
$ \bar{\nu^c}_{\ell' L} \nu_{\ell L}$ is left out 
assuming that the corresponding
coefficients are vanishing or negligibly small.)

The first term is a Dirac mass term, analogous to the mass term of charged
leptons. It conserves the total lepton number, but might violate the 
individual lepton flavor numbers. 

The second term is a Majorana mass term which breaks the total lepton number
conservation by two units. It is allowed only if the neutrinos have
no additive conserved charges of any kind.

If there are 3 left handed neutrinos $\nu_{\ell L}$, and $n$ additional
sterile neutrinos $\nu_{jR}$ it is convenient to define 
\begin{equation}
\nu = \left( \begin{array}{c} 
\nu_{\ell L} \\ (\nu_{jR})^c \end{array} \right)~,~~ 
- L_M = \frac{1}{2} \bar{\nu^c}M_{\nu}\nu +   {\rm H.c.} ~,
~~ M_{\nu} = \left( \begin{array}{cc} 
0 & {\mathcal M}^D \\ ({\mathcal M}^D)^T & {\mathcal M}^M 
\end{array} \right) ~.
\label{E;massp}
\end{equation}
The matrix $M_{\nu}$ is a symmetric complex matrix
and $\mathcal M^T$ is the transposed matrix. After diagonalization 
it has $3+n$ mass eigenstates $\nu_k$ that represent Majorana
neutrinos $(\nu_k^c = \nu_k)$.

From the point of view of phenomenology, there are several 
cases to be discussed. The seesaw mechanism corresponds to the case
when the scale of ${\mathcal M}^M$ is very large. There are three light
active Majorana neutrinos in that case.

On the other hand, if the scale of  ${\mathcal M}^M$ is not too high
when compared to the electroweak symmetry breaking scale,
there could be more than three light Majorana neutrinos,  mixtures of active
and sterile. Finally, when ${\mathcal M}^M = 0$ there are six 
massive Majorana neutrinos that merge to form three massive Dirac
neutrinos. The unitary matrix diagonalizing the mass term is
a $3 \times 3$ matrix in that case. In these latter cases
the `natural' explanation of the lightness of neutrinos is missing.

We shall not speculate further on the pattern of the mass matrix,
even though a vast literature exists on that subject
(see e.g. \cite{BMW} for a partial list of recent references).
In the interesting case of $N$ light left-handed  Majorana neutrinos,
the mass matrix is determined by $N$ neutrino masses, $N(N-1)/2$ mixing
angles, $(N-1)(N-2)/2$ CP violating phases  common to Dirac
and Majorana neutrinos, and $(N - 1)$ Majorana phases that affect
only processes which violate the total lepton number
(for the discussion of these Majorana phases see e.g. \cite{GKM}).

\subsection{Neutrino oscillation in vacuum}

As stated earlier, the neutrinos participating in the charged current 
weak interactions (the usual way neutrinos are observed
and the only way their flavor can be discerned) are characterized
by the flavor ($e, \mu, \tau$). But the neutrinos of definite flavor
are not necessarily states of a definite mass. Instead, they are generally
coherent superpositions of such states,
\begin{equation}
| \nu_{\ell} \rangle = \sum_i U_{\ell i} | \nu_i \rangle ~.
\label{e:super}
\end{equation}
When the standard model is extended to include neutrino mass, the 
mixing matrix $U$ is unitary.

In vacuum, the mass eigenstates propagate as plane waves. Leaving out 
the common phase, a beam of ultrarelativistic 
neutrinos $ | \nu_i \rangle $ with energy
$E$ at the distance $L$ acquires a phase
\begin{equation}
 | \nu_i (L) \rangle \sim  | \nu_i (L=0) \rangle 
\exp (-i\frac{m_i^2}{2}\frac{L}{E})
\label{e:phase}
\end{equation}
Given that, the amplitude of the process $\nu_{\ell} \rightarrow \nu_{\ell'}$
is 
\begin{equation}
A(\nu_{\ell} \rightarrow \nu_{\ell'}) = \sum_i U_{\ell i}e^{-i\frac{m_i^2 L}{2E}}
U_{\ell'i}^* ~,
\label{e:ampl}
\end{equation}
and the probability of the flavor change for $\ell \ne \ell'$
is  the square of this amplitude.
It is obvious that due to the unitarity of $U$ there is no flavor
change if all masses vanish or are exactly degenerate. 
The existence of the oscillations is a simple consequence of the coherence
in Eq. (\ref{e:super}). The detailed description of the quantum mechanics of
the oscillations in terms of wave packets is subtle and a bit involved
(see, e.g. the reviews \cite{packets} and references therein).
It includes, among ather things, the concept of coherence length, a distance
after which there is no longer coherence in Eq. (\ref{e:super}). For the purpose of this
review such issues are irrelevant, and will not be discussed further.  

The idea of oscillations was discussed early on  
by Pontecorvo \cite{Pont1,Pont2}
and by Maki, Nakagawa and Sakata \cite{MNS}. Hence, the mixing matrix $U$
is often associated with these names and the notation $U_{MNS}$ or $U_{PMNS}$
is used.

The formula for the probability is  particularly simple when only two
neutrino flavors, $\nu_{\ell}$ and  $\nu_{\ell'}$, mix appreciably,
since only one mixing angle is then relevant, 
\begin{equation}
P(\nu_{\ell} \rightarrow \nu_{\ell' \ne \ell}) =
\sin^22\theta \sin^2 \left[ 1.27 \Delta m^2 ({\rm eV^2}) 
\frac{L({\rm km})}{E_{\nu}({\rm GeV})} \right] ~,
\label{e:osc2}
\end{equation}
where the appropriate factors of $\hbar$ and $c$ were included
(the same is obtained when the length is in meters and energy 
in MeV). Here $\Delta m^2 \equiv |m_2^2 - m_1^2|$ (note that
the sign is irrelevant) is the mass squared difference.

Thus, the oscillations in this simple case
are characterized by the oscillation length
\begin{equation}
L_{osc}({\rm km})  = \frac{2.48 E_{\nu} {\rm (GeV)}}{\Delta m^2 {\rm (eV^2)}} ~,
\label{e:length}
\end{equation}
and by the amplitude $\sin^22\theta$. 

For obvious reasons the oscillation studies are optimally performed at distances
$L \sim L_{osc}$ from the neutrino source. At shorter distances the oscillation
amplitude is reduced and at larger distances the neutrino flux is reduced making
the experiment more difficult. Note that at distance $L \gg L_{osc}$ the oscillation
pattern is smeared out and the oscillation probability (\ref{e:osc2}) approaches
$(\sin^22\theta/2)$ and becomes independent of $\Delta m^2$.

The mixing matrix of 3 neutrinos is parametrized by three angles, 
conventionally denoted as
$\theta_{12}, \theta_{13},  \theta_{23}$,  one $CP$ violating phase $\delta$
and two Majorana phases $\alpha_1, \alpha_2$. Using $c$ for the cosine and
$s$ for the sine, we write $U$ as

\begin{eqnarray}
\hspace{-1.6cm} \left( \begin{array}{c}
\nu_e \\ \nu_{\mu} \\ \nu_{\tau}
\end{array} \right)
  =  
 \left( \begin{array}{ccc}
c_{12}c_{13} & s_{12}c_{13} & s_{13}e^{-i\delta} \\
-s_{12}c_{23}-c_{12}s_{23}s_{13}e^{i\delta} & 
c_{12}c_{23}-s_{12}s_{23}s_{13}e^{i\delta} &
s_{23}c_{13} \\
s_{12}s_{23}-c_{12}c_{23}s_{13}e^{i\delta} & 
-c_{12}s_{23}-s_{12}c_{23}s_{13}e^{i\delta} &
c_{23}c_{13}
\end{array} \right)
\left( \begin{array}{r}
e^{i\alpha_1/2}~\nu_1 \\ e^{i\alpha_2/2}~\nu_2 \\ \nu_3
\end{array} \right)  ~.
\label{e:u3}
\end{eqnarray}
 
By convention the mixing angle $\theta_{12}$ is associated with the solar neutrino
oscillations, hence the masses $m_1$ and $m_2$ are separated by the smaller interval
$\Delta m_{sol}^2$ 
(we shall assume, again by convention, that $m_2 > m_1$)
while $m_3$ is separated from the 1,2 pair by the larger interval
$\Delta m_{atm}^2$, and can be either lighter or heavier than  $m_1$ and $m_2$.
The situation where $m_3 > m_2$  is called `normal hierarchy', while
the `inverse hierarchy' has $m_3 < m_1$.
Not everybody follows these conventions, so caution should be used when comparing
the various results appearing in the literature.

The general formula for the probability that the ``transition''
$\ell \rightarrow \ell'$ happens at $L$ is
\begin{eqnarray}
P(\nu_{\ell} \rightarrow \nu_{\ell'}) & = &
| \sum_i U_{\ell i} U_{\ell'i}^* e^{-i(m_i^2/2E)L} |^2 \nonumber \\
& = & \sum_i | U_{\ell i} U_{\ell' i}^*|^2
+ \Re \sum_i \sum_{j \ne i} U_{\ell i}U_{\ell' i}^*U_{\ell j}^*U_{\ell' j}
e^{i\frac{|m_i^2 - m_j^2|L}{2E}} ~.
\label{e:oscg}
\end{eqnarray}

Clearly, the probability (\ref{e:oscg}) is independent of the Majorana phases
$\alpha$. The oscillations described by the Eq.(\ref{e:oscg}) violate 
the individual flavor lepton numbers, but conserve the total lepton number.
The oscillation pattern is identical for Dirac or Majorana neutrinos.

The general formula
can be simplified in several cases of practical importance.
For 3 neutrino flavors, using the empirical fact that
$\Delta m_{atm}^2 \gg \Delta m_{sol}^2$ and considering distances comparable
to the atmospheric neutrino oscillation length, only three parameters are
relevant in the zeroth order, 
the angles $\theta_{23}$ and  $\theta_{13}$ and 
$\Delta_{atm} \equiv \Delta m_{atm}^2L/4E_{\nu}$.
However, corrections of the first order in 
$\Delta_{sol} \equiv \Delta m_{sol}^2L/4E_{\nu}$
should be also considered and are included below
(some of the terms with $\Delta_{sol}$ are further reduced by the presence
of the emipirically small $\sin^22\theta_{13}$):
\begin{eqnarray}
P(\nu_{\mu} \rightarrow \nu_{\tau}) & \simeq & \cos^4\theta_{13}
\sin^22\theta_{23} \sin^2 \Delta_{atm}  \\
& - & \Delta_{sol} \cos^2 \theta_{13} \sin^2 2\theta_{23} 
(\cos^2\theta_{12} - \sin^2 \theta_{13}\sin^2\theta_{12}) 
\sin 2 \Delta_{atm} \nonumber \\
& - & \Delta_{sol} \cos \delta \cos \theta_{13}
\sin2\theta_{12} \sin2\theta_{13} \sin2\theta_{23}  \cos2\theta_{23}
\sin 2 \Delta_{atm} /2
\nonumber \\ 
& + &  \Delta_{sol} \sin\delta
\cos\theta_{13} \sin2\theta_{12} \sin2\theta_{13} \sin2\theta_{23} 
\sin^2 \Delta_{atm} ~, \nonumber
\label{e:oscs1}
\end{eqnarray}
\begin{eqnarray}
P(\nu_{\mu} \rightarrow \nu_e) & \simeq  & \sin^22\theta_{13}\sin^2\theta_{23}
\sin^2 \Delta_{atm} \\
& - & \Delta_{sol} \sin^2\theta_{23}\sin^2\theta_{12} \sin^22\theta_{13}
\sin 2\Delta_{atm} \nonumber \\ 
& + & \Delta_{sol} \cos \delta
\cos\theta_{13}\sin2\theta_{13}\sin2\theta_{23}\sin2\theta_{12}
 \sin2\Delta_{atm}/2 \nonumber \\
& - &  \Delta_{sol} \sin\delta
\cos\theta_{13} \sin2\theta_{12} \sin2\theta_{13} \sin2\theta_{23} 
\sin^2 \Delta_{atm} ~, \nonumber 
\label{e:oscs2}
\end{eqnarray}
\begin{equation}
P(\nu_{\mu} \rightarrow \nu_{\mu}) = 1 - P(\nu_{\mu} \rightarrow \nu_e)
- P(\nu_{\mu} \rightarrow \nu_{\tau}) ~,
\end{equation}
where $\delta$ is the $CP$ phase of Eq. (\ref{e:u3}) and
\begin{eqnarray}
P(\nu_e \rightarrow \nu_x) & \simeq &
\sin^22\theta_{13} \left[ \sin^2 \Delta_{atm}
 -  \Delta_{sol} \sin^2\theta_{12} 
\sin 2\Delta_{atm} \right] \nonumber \\
& + & 
\Delta^2_{sol} \cos^4\theta_{13}\sin^22\theta_{12} ~, 
\label{e:oscs3}
\end{eqnarray}
where the term linear in $\Delta_{sol}$ is suppressed by
the factor  $\sin^22\theta_{13}$ and therefore the quadratic
term is also included. The $\nu_e$ disappearance probability
is independent of the $CP$ phase $\delta$.

Note that the terms of first order in $\Delta_{sol}$ depend
on the sign of $\Delta_{atm}$, i.e., on the hierarchy.

On the other hand, at distances much larger than  
the atmospheric neutrino oscillation length, the 
electron neutrino and antineutrino disappearance is governed by
\begin{equation}
P(\nu_e \rightarrow \nu_x) \simeq 1  - \sin^4\theta_{13} 
- \cos^4\theta_{13} \left[ 1 - \sin^22\theta_{12} \sin^2 \Delta_{sol} \right] ~,
\label{e:oscs4}
\end{equation}
In both of the latter cases $\nu_x$ denotes any neutrino except $\nu_e$.
(General expressions for mixing of three neutrinos 
in vacuum can be found
in Eqs. (7)-(12) of Ref. \cite{BMW}.) 

In the mixing matrix (\ref{e:u3}) the presence of the phase $\delta$ signifies
the possibility of $CP$ violation, the expectation that
\begin{equation}
P(\nu_{\ell'} \rightarrow \nu_{\ell})
\neq P(\bar{\nu}_{\ell'} \rightarrow \bar{\nu}_{\ell})~,
\label{e:cp}
\end{equation}
i.e., that for example the probability of $\nu_{\mu}$ oscillating
into $\nu_e$ is different from the probability
of $\bar{\nu}_{\mu}$ oscillating into $\bar{\nu}_e$.

The magnitude of the $T$ or $CP$ violation is characterized
by the differences
\begin{eqnarray}
& & P(\bar{\nu}_{\mu}\rightarrow\bar{\nu}_e)-P(\nu_{\mu} \rightarrow \nu_e) =
-[P(\bar{\nu}_{\mu} \rightarrow \bar{\nu}_{\tau})-
P(\nu_{\mu} \rightarrow \nu_{\tau})]
\nonumber  \\
& = & P(\nu_e \rightarrow \nu_{\tau})-
P(\bar{\nu}_e \rightarrow \bar{\nu}_{\tau}) \\
& = &
-4c_{13}^2s_{13}c_{23}s_{23}c_{12}s_{12}\sin\delta[\sin2\Delta_{12} + 
\sin2\Delta_{23} +  \sin2\Delta_{31}]  \nonumber  \\
& = & 16c_{13}^2s_{13}c_{23}s_{23}c_{12}s_{12}
\sin\Delta_{12}\sin\Delta_{23} \sin\Delta_{31} ~, \nonumber
\label{e:cpf}
\end{eqnarray}
where, as before, $\Delta_{ij} = (m_i^2 - m_j^2) \times L /4E$.
Thus, the size of the effect is the same in all three channels,
and $CP$ violation is observable
only if all three masses are different (i.e., nondegenerate),
and all three angles are nonvanishing.
The possibility of $CP$ violation in the lepton sector was first discussed
in \cite{Cab78,BWP80}.

\subsection{Neutrino oscillations in matter}

The oscillation phenomenon has its origin in the phase difference between the
coherent components of the neutrino flavor 
eigenstates described by Eq. (\ref{e:phase}).
When neutrinos propagate in matter  additional contributions
to the phase appear, besides the one caused by the nonvanishing 
mass of the state $\nu_i$.
To see the origin of such phase, consider the effective Hamiltonian of neutrinos
in presence of matter. Obviously, only  phase differences are of importance.

Without invoking any nonstandard interactions two effects are present.
All active neutrinos interact with quarks and electrons by the neutral current
weak interactions ($Z$ exchange), but only electron neutrinos and antineutrinos
interact with electrons by the exchange of $W$. The corresponding effective
potential is \footnote{The effective potential was introduced first by Wolfenstein
\cite{Wol78}, and used also in Refs. \cite{Bar80,Lew80} who corrected the missing
$\sqrt{2}$ in the original paper. Finally, the correct sign was obtained in
Ref.\cite{Lan83}.} ($V_C$ stands for the charged current)
\begin{equation}
V_C(\nu_e) = \sqrt{2}G_F N_e ~,~~{\rm and}~  V_C(\bar{\nu}_e) = -\sqrt{2}G_F N_e~,
\label{e:matpot}
\end{equation}
where $N_e$ is the electron number density. (There are no $\mu$ or $\tau$ leptons in
normal matter, hence there is no analogous potential for $\nu_{\mu}$ or
$\nu_{\tau}$.) In practical units
\begin{equation}
V_C = 7.6 Y_e \frac{\rho}{10^{14}{\rm (g/cm^3)}} {\rm (eV)}, 
~~Y_e = \frac{N_e}{N_p + N_n} ~. 
\end{equation}

Similarly, any active neutrino acquires an effective potential due
to the neutral-current interaction $V_N = - G_F N_n/\sqrt{2}$.
The corresponding effective potential is absent for sterile
neutrinos.

Given the effective potential, Eq.(\ref{e:matpot}), electron neutrinos
travelling distance $L$ in matter of constant density $N_e$
acquire an additional phase
\begin{equation}
\nu_e (L) = \nu_e (0) e^{-i\sqrt{2}G_F N_e L} ~.
\end{equation}
The corresponding matter oscillation length \cite{Wol78} is therefore
\begin{equation}
L_0 = \frac{2 \pi}{\sqrt{2}G_F N_e} ~~
\simeq \frac{ 1.7 \times 10^7 {\rm (m)}}{\rho {\rm (g/cm^3)}~ Y_e} ~.
\label{e:matlength}
\end{equation}
Unlike the vacuum oscillation length, Eq.(\ref{e:length}), the matter 
oscillation length $L_0$ is independent of the neutrino energy.
Note that the matter oscillation length in rock is $L_0 \approx 10^4$ km,
and in the center of the Sun $L_0 \approx$ 200 km.

Considering for simplicity just two mass eigenstates $\nu_1$ and $\nu_2$
that are components of the flavor eigenstates $\nu_e$ and $\nu_{\alpha}$
with the mixing angle $\theta$,
we obtain the  time (or space) development Schr\"{o}dinger equation
\begin{eqnarray} 
i \frac{d}{dt} \left( \begin{array}{c}  \nu_1 \\ \nu_2 \end{array}  \right) =
 \left( \begin{array}{cc} \frac{m_1^2}{2E} + V_C c^2 & ~~~~V_C sc \\
V_C sc & ~~~~\frac{m_2^2}{2E} + V_C s^2 \end{array} \right)
 \left( \begin{array}{c}  \nu_1 \\ \nu_2  \end{array} \right)  ~,
\label{e:sch_mat}
\end{eqnarray} 
where, as before $c = \cos\theta$ and $s = \sin\theta$.

The $2 \times 2$ matrix above can be brought to the diagonal form
by the transformation
\begin{eqnarray}
\nu_{1m} & =  & \nu_e \cos\theta_m - \nu_{\alpha} \sin \theta_m \\ \nonumber
\nu_{2m} & =  & \nu_e \sin\theta_m + \nu_{\alpha} \cos \theta_m ~,
\end{eqnarray}
where the new mixing angle in matter, $\theta_m$, depends on the vacuum mixing angle 
$\theta$ and on the vacuum and matter oscillation lengths $L_{osc}$ and $L_0$,
\begin{equation}
\tan 2\theta_m = 
\tan 2\theta \left( 1 - \frac{L_{osc}}{L_0 \cos 2\theta} \right)^{-1} ~.
\end{equation}
The effective oscillation length in matter is then
\begin{equation}
L_m = L_{osc} \frac{\sin 2\theta_m}{\sin 2\theta} =
 L_{osc} \left[ 1 + \left( \frac{L_{osc}}{L_0} \right)^2
- \frac{2 L_{osc}}{L_0} \cos 2\theta \right]^{-1/2} ~,
\end{equation}
and the probability of detecting $\nu_e$ at a distance $L$ from the
$\nu_e$ source has the usual form, but with $\theta \rightarrow \theta_m$
and $L_{osc} \rightarrow L_m$,
\begin{equation}
P(E_{\nu},L,\theta,\Delta m^2) = 1 - \sin^2 2\theta_m \sin^2 \frac{\pi L}{L_m} ~.
\end{equation}

When considering oscillations with two flavors, the mixing angle $\theta$
can be restricted to the interval $(0,\pi/2)$. In vacuum, only $\sin^2 2\theta$
is relevant, and hence only half of that interval,  $(0,\pi/4)$ could be used.
However, once matter oscillations are present, also the $\cos 2\theta$ 
becomes relevant, and thus the whole $(0,\pi/2)$ interval might be needed.
The part of the parameter space corresponding to the $(\pi/4,\pi/2)$ angles
was called the ``dark'' side in Ref.\cite{Mur00}, who suggested using
$\tan^2 \theta$ in the plots instead of  $\sin^2 2\theta$. (The advantage
of  $\tan^2 \theta$ is that when plotted on the log scale, the reflection 
symmetry around  $\tan^2 \theta =1$ for vacuum oscillations is maintained.)
The importance of allowing the full range of the mixing angles in the
three flavor analysis was stressed earlier, see e.g. Ref. \cite{Fog96}.

With our convention $m_2 > m_1$ mixing angles in the interval 
 $(\pi/4,\pi/2)$ would mean that $\nu_e$ contains dominantly
the heavier component $\nu_2$. (This is not the case in practice, see
Sect. {\it 3.1}.)
It appears now that at least two of the three mixing angle are large
(and $\le \pi/4$),
and the constraint on the third one, $\theta_{13}$ are still such that linear
plots are more revealing. Therefore, in the following we use the traditional
plots with  $\sin^2 2\theta$.

The same results can be obtained, naturally, by rewriting the equation of motion
in the flavor basis
\begin{eqnarray} 
i \frac{d}{dx} \left( \begin{array}{c}  \nu_e \\ \nu_{\alpha} \end{array}  \right) =
 2 \pi \left( \begin{array}{cc} - \frac{\cos2\theta}{L_{osc}} + \frac{1}{L_0} &
\hspace{0.8cm} \frac{\sin2\theta}{2 L_{osc}}  \\ 
 \frac{\sin2\theta}{2 L_{osc}} & \hspace{0.8cm} 0 \end{array} \right)
 \left( \begin{array}{c}  \nu_e \\ \nu_{\alpha}  \end{array} \right)  ~,
\label{e:sch_fl}
\end{eqnarray} 
Here one can see clearly that in matter,
unlike in vacuum, the oscillation pattern depends on whether the mixing
angle $\theta$ is smaller or larger than $\pi/4$. 
(For antineutrinos the sign in front of $1/L_0$ is reversed.)

In matter of a constant density we can now consider several special cases:
\begin{itemize}
\item Low density limit, $L_{osc} \ll L_0$. In this case matter has a rather small
effect. On Earth, one is able to observe oscillations 
only provided $L_{osc} < $ Earth
diameter, and therefore this limit applies. 
The matter effects could be perhaps observed 
as small day-night variation in the solar neutrino signal. 
In long baseline oscillation
experiments matter effects could cause difference in the oscillation probability of
neutrinos and antineutrinos, hence these effects, together with
the hierarchy problem (i.e., whether $m_3 > m_2$ or $m_3 < m_1$) 
are crucial in the search for $CP$
violation and its interpretation.
\item High density limit, $L_{osc} \gg L_0$. The oscillation amplitude is suppressed
by the factor $L_0/|L_{osc}|$.  For $m_2^2 > m_1^2$ the matter mixing angle is
$\theta_m \rightarrow 90^0$ and thus $\nu_e \rightarrow \nu_2$.
\item $|L_{osc}| \approx L_0$. In this case the matter effects can be enhanced.
In particular, for $L_{osc}/L_0 = \cos2\theta$ one has $\sin^2 2\theta_m = 1$,
i.e. the maximum mixing, even
for small vacuum mixing angle $\theta$. This is the basis of the 
Mikheyev-Smirnov-Wolfenstein (MSW) effect \cite{Wol78,MS86}.
\end{itemize}

When neutrinos propagate in matter of varying density, the equations of motion
 (\ref{e:sch_mat}) or (\ref{e:sch_fl})  must be solved. There is a vast literature
on the subject, in particular on the application to the solar neutrinos.
There, the $\nu_e$ are often created at densities above the resonance density,
and therefore are dominantly in the higher mass eigenstate. When the
resonance density is reached, the two instantaneous eigenvalues are almost
degenerate. In the adiabatic regime, the neutrino will remain in
the upper eigenstate, and when it reaches vacuum it will be  
(for small vacuum mixing angle $\theta$)
dominantly in the state that we denoted
above as $\nu_{\alpha}$. In the general case, there is a finite probability
$P_x$ for jumping from one eigenstate to the other one, and the
conversion might be incomplete.
The average survival probability is \cite{parke86}
\begin{equation}
\langle P(\nu_e \rightarrow \nu_e) \rangle = \frac{1}{2}
[1 + (1 - 2P_x)\cos 2\theta_m(\rho_{max}) \cos 2\theta ] ~,
\end{equation}
Usually $\cos 2\theta_m(\rho_{max}) \simeq -1$ and thus
$\langle P(\nu_e \rightarrow \nu_e) \rangle \simeq
\sin^2 \theta + P_x\cos 2\theta$.

The transition point between the regime of vacuum and matter oscillations is
determined by the ratio
\begin{equation}
\frac{L_{osc}}{L_0} = \frac{2\sqrt{2} G_F N_e E_{\nu}}{\Delta m^2} = 
0.22 \left[\frac{E_{\nu}}{{\rm 1~ MeV}} \right] 
\left[\frac{\rho Y_e}{{\rm 100 g/cm^3}} \right]
\left[ \frac{7 \times 10^{-5}{\rm eV^2}}{\Delta m^2} \right]
\label{e:lrat}
\end{equation}
If this fraction is larger than unity, the matter oscillations dominate, and
when this ratio is less than $\cos 2\theta$ the vacuum oscillations dominate.
Generally, there is a smooth transition in between these two regimes.

\begin{figure}[htb]
\begin{center}
\centerline{\includegraphics[width=3.5 in]{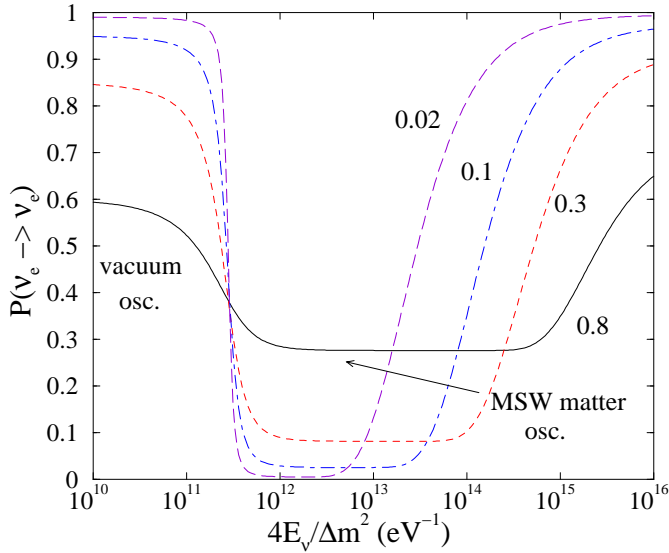}}
\bigskip
\caption{ Schematic illustration of the survival probability of $\nu_e$ created
at the solar center. The curves are labelled by the $\sin^2 2\theta$ values.  
}
\label{fig:surv}
\end{center} 
\end{figure}

The electron neutrino survival probability is illustrated in Fig. \ref{fig:surv},
where it is plotted against $4E_{\nu}/\Delta m^2$, see Eq.(\ref{e:lrat}).
In Fig. \ref{fig:surv} it was assumed that 
the neutrinos originated in the center of the Sun,
hence the relatively sharp feature at 
$4E_{\nu}/\Delta m^2 \sim 2.6 \times 10^{11}$ eV$^{-1}$
where according to the Eq.((\ref{e:lrat}) $L_{osc}/L_0 = 1$ for the central density
of the Sun. 
Note that below and above this dividing line the survival probability is 
almost independent
of the neutrino energy, hence  no spectrum distortion is expected. 
The subsequent increase in the neutrino survival probability
for larger values of $4E_{\nu}/\Delta m^2$ is caused by the `nonadiabatic'
transition, i.e. by the gradual increase of the jump probability
towards the limiting value $P_x \rightarrow 1$.

For the parameters
corresponding to the preferred solar solution ($\sin^2 2\theta \simeq 0.83$
and $\Delta m^2 \sim 7\times 10^{-5}$ eV$^2$) the $pp$ neutrinos
with $4E_{\nu}/\Delta m^2 < 2.4\times 10^{10}$ eV$^{-1}$ and the $^7$Be
neutrinos with $4E_{\nu}/\Delta m^2 = 4.9\times 10^{10}$ eV$^{-1}$
undergo vacuum oscillations, while the $^8$B neutrinos
with  $4E_{\nu}/\Delta m^2 >  2.55\times 10^{11}$ eV$^{-1}$ undergo
MSW matter oscillations. (Clearly, that must be the case since otherwise it would
be impossible to understand the $\sim$ 0.3 suppression of the $^8$B neutrino $\nu_e$
flux observed in the charged current reactions, see {\it 3.1}.)

\subsection{Tests of CP, T and CPT invariance}

In vacuum $CP$ conservation implies that 
$P(\nu_{\ell} \rightarrow \nu_{\ell'}) = 
P(\bar{\nu}_{\ell} \rightarrow \bar{\nu}_{\ell'})$ (see Eq.(\ref{e:cpf}).
Violation of this inequality thus would mean that $CP$ is not conserved in the lepton
sector. Substantial effort is devoted to the $CP$ tests. 
Matter effects can induce inequality
between $P(\nu_{\ell} \rightarrow \nu_{\ell'})$ and 
$P(\bar{\nu}_{\ell} \rightarrow \bar{\nu}_{\ell}')$
and so the analysis must carefully account for them.

If $CP$ is not conserved, but $CPT$ invariance holds, then 
$T$ invariance will be also violated, i.e. 
$P(\nu_{\ell} \rightarrow \nu_{\ell'}) - P(\nu_{\ell'} \rightarrow \nu_{\ell})$ 
need not vanish. 
Here matter effects cannot mimic the apparent $T$ invariance violation.

Finally, if $CPT$ is not conserved, then  
$P(\nu_{\ell} \rightarrow \nu_{\ell'}) - 
P(\bar{\nu}_{\ell'} \rightarrow \bar{\nu}_{\ell})$ 
might be nonvanishing. Many tests of the $CPT$ invariance in the neutrino
sector have been suggested, see e.g. \cite{BPWW} or references listed in
\cite{BMW}.

$CPT$ invariance is based on Lorentz invariance, hermiticity of the Hamiltonian 
and locality. Its violation would have, naturally, enormous consequences.
Yet there are many proposed scenarios of $CPT$ violation, in particular in 
the neutrino sector (for a whole series of papers on that topic see e.g.
\cite{Kos03}).   

$CPT$ invariance implies that neutrino and antineutrino masses are equal. 
If that is not true, then the $\Delta m^2$ as determined
in the solar neutrino experiments (thus involving $\nu_e$)
might not be the same as the  $\Delta m^2$
needed to explain the LSND result which involves 
$\bar{\nu}_{\mu} \rightarrow \bar{\nu}_e$ oscillations,
see {\it 3.1.4} below. That was the gist
of the phenomenological proposal in Ref. \cite{MY01}. (See also further
elaboration in \cite{Bar02}.)

With the demonstration of the consistency between the observed solar $\nu_e$ deficit
and the disappearance of reactor $\bar{\nu}_e$ 
by the KamLAND collaboration (see {\it 3.1.3} below),
this possibility seems unlikely, even though a proposal
has been made to accommodate $CPT$ violation in that context 
\cite{Bar03}, see also \cite{GMS03}.  
As has been shown in \cite{Mur03}, the consistency
of the solar $\nu$ oscillation solution and 
the KamLAND reactor result can be interpreted
as a test of $CPT$ giving
\begin{equation}
|\Delta m_{\nu}^2 - \Delta m_{\bar{\nu}}^2| < 1.3 \times 10^{-3} {\rm ~eV^2} ~~
90\% {\rm ~CL} ~,
\end{equation}
where $\Delta m^2_\nu$ and $\Delta m^2_{\bar \nu}$ refer to the mass eigenstates
$\nu_1$ and $\nu_2$ involved in the observed solar and reactor neutrino oscillations.

To test for the $CP$ invariance violation
experimentally, one would compare the probabilities
$P(\nu_{\mu} \rightarrow \nu_e)$ and 
$P(\bar{\nu}_{\mu} \rightarrow \bar{\nu}_e)$. This could
be done realistically with $\sim$ 1 GeV $\nu_{\mu}$ beams at a distance 
$L \sim E_{\nu}/\Delta m_{atm}^2$ such that the contribution involving $\Delta m_{sol}^2$
are small. The effect of matter, however, must be included.
Using the notation
\begin{equation}
s_{ij} = \sin\theta_{ij},~c_{ij} = \cos\theta_{ij},
~\Delta_{ij} = \Delta m_{ij}^2 L/4E_{\nu}
\end{equation}
we obtain the formula
\begin{eqnarray}
P(\nu_{\mu} & \rightarrow  & \nu_e)   =  
4 c_{13}^2 s_{13}^2 s_{23}^2 \sin^2 \Delta_{31}
\hspace{7cm} \nonumber \\
 & + & 8 c_{13}^2 s_{13} s_{23} c_{23} s_{12} c_{12}\sin\Delta_{31}
[ \cos\Delta_{32}\cos\delta  -   \sin\Delta_{32}\sin\delta ]  \sin\Delta_{21}
\nonumber \\
& - & 8 c_{13}^2 s_{13}^2 s_{23}^2 s_{12}^2 
\cos\Delta_{32}\sin\Delta_{31}\sin\Delta_{21}
\nonumber \\
& + & 4 c_{13}^2 s_{12}^2 [ c_{12}^2 c_{23}^2  +  s_{12}^2 s_{23}^2 s_{13}^2
 - 2 c_{12}  c_{23} s_{12} s_{23} s_{13} \cos\delta ] \sin^2 \Delta_{21}
\nonumber \\
& - & 8  c_{13}^2 s_{13}^2 s_{23}^2 (1  -  2 s_{13}^2) \frac{a L}{4 E_{\nu}}
\sin\Delta_{31} \left[\cos\Delta_{32} - \frac{\sin\Delta_{31}}{ \Delta_{31}}
\right] ~.
\label{e:longb}
\end{eqnarray}
Here the first term gives the largest effect, while the terms in the third and fourth
line represent small $CP$ conserving corrections (proportional to $\sin\Delta_{21}$
and $\sin^2\Delta_{21}$.  The term with $\sin\delta$ in the second line violates
$CP$ symmetry, while the term with $\cos\delta$ preserves it. Finally, the term with 
$a L/4 E_{\nu}$ in the last line represents the matter effects. 

The matter effects are characterized by
\begin{equation}
\hspace{-0.5cm} a = 2 \sqrt{2} G_F N_e E_{\nu} = 
1.54 \times 10^{-4} Y_e \rho( {\rm g/cm^3} ) 
E_{\nu} ({\rm GeV}) ~ ( a~ {\rm is~ in~ (eV^2))}.
\end{equation}
The probability $P(\bar{\nu}_{\mu} \rightarrow \bar{\nu}_e)$ is obtained by the
substitution $\delta \rightarrow -\delta$ and $a \rightarrow -a$.

To test for $CP$ symmetry, one would determine
\begin{eqnarray}
A_{CP}   & = &
\frac{P(\nu_{\mu} \rightarrow \nu_e) - P(\bar{\nu}_{\mu} \rightarrow \bar{\nu}_e)}
{P(\nu_{\mu} \rightarrow \nu_e) + P(\bar{\nu}_{\mu} \rightarrow \bar{\nu}_e)}
\nonumber \\
&  \simeq &  - \Delta_{21} \frac{\sin 2 \theta_{12}}{\sin \theta_{13}}
\sin\delta - \frac{ a L}{2 E_{\nu}} \frac{ \cos \Delta_{32}}{ \sin \Delta_{31}} ~,
\label{e:acp}
\end{eqnarray}
where we used the empirical fact that $\cos \theta_{23} \sim \sin \theta_{23}$
and $\sin(\Delta_{21}) \sim \Delta_{21}$
for the distances and energies usually considered. 
Since $\theta_{13}$ is small, the $CP$ asymmetry can be enhanced.
However, the individual terms in Eq. (\ref{e:acp}) depend on  $\theta_{13}$,
so for smaller  $\theta_{13}$ it is more difficult to reach the required
statistical precision.

There are several 
parameter degeneracies in Eq. (\ref{e:acp}) when separate measurements
of $P(\nu_{\mu} \rightarrow \nu_e)$ and  $P(\bar{\nu}_{\mu} \rightarrow \bar{\nu}_e)$
are made at given $L$ and $E_{\nu}$. 
($i$) There can be two values $\delta , \theta_{13}$
and $\delta' , \theta_{13}'$ leading to the same probabilities. $(ii)$ 
Sign of $\Delta_{31}$ and $\Delta_{32}$
(i.e. the normal or inverted hierarchy), 
where one set  $\delta , \theta_{13}$ gives
the same oscillation probabilities with one sign, 
as another set   $\delta' , \theta_{13}'$
with the opposite sign of $\Delta_{31}$.($iii$) Since the mixing angle 
$\theta_{23}$ is determined in experiments sensitive 
only to $\sin^2 2\theta_{23}$ there
is an ambiguity between $\theta_{23}$ and $\pi/2 - \theta_{23}$. However, for the
preferred value  $\theta_{23} \sim \pi/4$ 
this ambiguity is essentially irrelevant.   
Various strategies to overcome the parameter degeneracies have been proposed.
The choice of the neutrino energy $E_{\nu}$ (and whether a wide or narrow
beam is used) and the distance $L$ play essential role. Clearly, if some
of the so far unknown parameters (e.g. $\theta_{13}$ or the sign of the hierarchy)
could be determined independently, some of these ambiguities would be diminished.   
(For some of the suggestions how to overcome the parameter degeneracy see e.g.
\cite{BNL03,BMW02}.)

\subsection{Violation of the total lepton number conservation}

Neutrino oscillations described so far are insensitive to the transformation 
properties under charge conjugation, i.e., whether neutrinos are Dirac or Majorana
particles. However, if the mass eigenstates $\nu_i$ are Majorana particles,
then $\nu \rightarrow \bar{\nu}$ oscillations that violate the total lepton
number conservation are possible. There are two kinds of such processes.

Recall that in the standard Model charged current processes the 
neutrino $\nu_{\ell}$ produces the negatively charged lepton $\ell^-$ while
 antineutrino $\bar{\nu}_{\ell}$ produces the positively charged antilepton
$\ell^+$. The standard model also requires that in order to produce the
lepton $\ell^-$ the neutrino $\nu_{\ell}$ must have (almost) purely negative
helicity, and similarly for  $\ell^+$ the $\bar{\nu}_{\ell}$ must have
 (almost) purely positive helicity. The amplitudes for the `wrong', i.e.,
suppressed helicity component is only of the order $m_{\nu}/E_{\nu}$ and
therefore vanishes for massless neutrinos. But we know now that neutrinos
are massive, although light, and the therefore
`wrong' helicity states are present.

When neutrinos are Majorana particles, the total lepton number may not
be conserved. Thus, neutrinos $\nu_{\ell}$ born together with the leptons 
$\ell^+$ can create leptons $\ell^+$ (or even a 
different flavor $\ell'^+$) as long as the
helicity rules are obeyed. The amplitude of this process is of the
order $m_{\nu}/E_{\nu}$, thus small,
independent of the distance the neutrinos
travel. Therefore, this first kind of the   $\nu \rightarrow \bar{\nu}$
transformation should not be really called oscillations. When such a
transformation happens inside a nucleus, it leads to the process
of neutrinoless double beta decay discussed in more detail below.

For Majorana neutrinos, there are several nonstandard processes
involving helicity flip in which left-handed neutrinos $\nu_L$
are converted into right-handed (anti)neutrinos $\nu_R^c$. This can happen
for Majorana neutrinos with a transition magnetic moment $\mu_{ij}$.
In a transverse magnetic field $B_{\bot}$, $\nu_{iL}$ can be connected
to $\nu_{jR}^c$. However, the transition probability is proportional
to the small quantity $|\mu_{\nu} B|$ that vanishes for massless
neutrinos. 
There are also models that predict neutrino decay
involving a helicity-flip and a Majoron $\chi$ production.  Again,
the decay rate is expected to be small.
If such processes exist one expects, among other things,
that solar $\nu_e$ could be subdominantly converted into $\bar{\nu}_e$.
The recent limit on the solar $\bar{\nu}_e$ flux, expressed as a fraction
of the Standard Solar Model $^8$B $\nu_e$ flux is
$2.8 \times 10^{-3}$ \cite{Kamnuebar}. (See also \cite{Kamnuebar} for
references to some of the theoretical model expectations for such  
$\nu \rightarrow \bar{\nu}$ conversion.)

In addition, there could be
also transitions without helicity flip, which require that
both Dirac and Majorana mass terms are present. These second class
oscillations \cite{BP76} involve transitions $\nu_L \rightarrow \nu^c_L$,
i.e., the final neutrino is sterile and therefore unobservable. 
Estimates of the experimental observation possibilities 
of the neutrino-antineutrino
oscillations are not encouraging \cite{LW98}.

\begin{figure}[htb]
\begin{center}
\includegraphics[width=3.5 in]{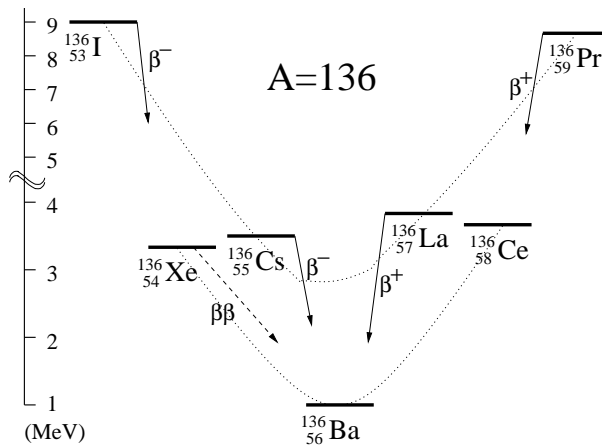}
\medskip
\caption{  Masses of nuclei with $A$ = 136.  The even-even and odd-odd nuclei
are connected by dotted lines. $^{136}$Xe is stable against 
ordinary $\beta$ decay,
but unstable against $\beta^-\beta^-$ decay. The same is true for $^{136}$Ce,
however, the  $\beta^+\beta^+$ decay is expected to be slower than the
 $\beta^-\beta^-$ decay.
}
\label{fig:parabola}
\end{center} 
\end{figure}

Study of the neutrinoless double beta decay ($0\nu\beta\beta$) appears
to be the best way to establish the Majorana nature of the neutrino,
and at the same time gain valuable information about the absolute 
scale of the neutrino masses.
Double beta decay is a rare transition between two nuclei with the
same mass number $A$ involving change of the nuclear charge $Z$
by two units. The decay can proceed only if the initial nucleus is
less bound than the final one, and both must be more bound than
the intermediate nucleus.
These conditions are fulfilled in nature
for many even-even nuclei, and only for them. Typically, the decay 
can proceed from the ground state (spin and parity always $0^+$) of the
initial nucleus to the ground state (also  $0^+$) of the final
nucleus, although the decay into  excited states
($0^+$ or $2^+$) is in some cases also energetically possible.
In Fig. \ref{fig:parabola} we show a typical situation for $A = 136$. 
There are 11 candidate nuclei (all for the  $\beta^-\beta^-$ decay)
with $Q$ value above 2 MeV, thus potentially useful for the study
of the $0\nu\beta\beta$ decay. (Large  $Q$ values are preferable
since the rate scales like $Q^5$ and the background suppression
is typically easier for larger $Q$.)

There are two modes of the double beta decay.
The two-neutrino decay, $2\nu\beta\beta$,
\begin{equation}
(Z,A)  \rightarrow (Z+2,A) + e_1^- + e_2^- + \bar{\nu}_{e1}  + \bar{\nu}_{e2}
\label{e:2nu}
\end{equation}
conserves not only electric charge but also
lepton number. On the other hand, the neutrinoless decay, 
\begin{equation}
(Z,A)  \rightarrow (Z+2,A) + e_1^- + e_2^-
\label{e:0nu}
\end{equation}
violates lepton number conservation. One can distinguish the two decay modes 
by the shape of the electron sum energy spectra, which 
are determined by the phase space of the outgoing light particles.
Since the nuclear masses are so much larger than the decay $Q$ value,
the nuclear recoil energy is negligible, and the electron sum energy
of the  $0\nu\beta\beta$ is simply a peak at $T_{e1} + T_{e2} = Q$
smeared only by the detector resolution.

The $2\nu\beta\beta$ decay is an allowed process with a very long lifetime 
$\sim 10^{20}$ years. It has been observed now in a number of cases
\cite{EV}. Observing the $2\nu\beta\beta$ decay is important 
not only as a proof
that the necessary background suppression has been achieved,
but also allows one to constrain the nuclear models needed
to evaluate the corresponding nuclear matrix elements.

The $0\nu\beta\beta$ decay involves a vertex changing two neutrons
into two protons with the emission of two electrons and nothing else.
One can visualize it by assuming that the process involves the exchange
of various virtual particles, e.g. light or heavy Majorana neutrinos,
right-handed current mediating $W_R$ boson, SUSY particles, etc.
No matter what the vertex is, the $0\nu\beta\beta$ decay
can proceed only when neutrinos are massive Majorana particles
\cite{SV82}. In the following we concentrate on the case
when the  $0\nu\beta\beta$ decay is mediated by the exchange
of light Majorana neutrinos interacting through the left-handed
$V - A$ weak currents. The decay rate is then,
\begin{equation}
[ T_{1/2}^{0\nu} (0^+ \rightarrow 0^+)]^{-1}
~=~ G^{0\nu}(E_0,Z) 
\left| M_{GT}^{0\nu} - \frac{g_V^2}{g_A^2} M_F^{0\nu} \right|^2
\langle m_{\beta\beta} \rangle^2 ~,
\label{e:0nut}
\end{equation}
where $G^{0\nu}$ is the exactly calculable phase space integral,
$\langle m_{\beta\beta} \rangle$ is the effective neutrino mass
and $M_{GT}^{0\nu}$,
$M_F^{0\nu}$ are the nuclear matrix elements.

The effective neutrino mass is 
\begin{equation}
 \langle m_{\beta\beta} \rangle = | \sum_i |U_{ei}|^2 m_{\nu_i} e^{i\alpha_i} | ~,
\label{e:meff}
\end{equation}
where the sum is only over light neutrinos ($m_i < 10$ MeV)\footnote{The
same quantity is sometimes denoted as 
$\langle m_{\nu} \rangle$ or $m_{ee}$}. The Majorana
phases $\alpha_i$ were defined earlier in Eq.(\ref{e:u3}). If the neutrinos
$\nu_i$ are $CP$ eigenstates, $\alpha_i$ is either 0 or $\pi$. Due to the presence
of these unknown phases, cancellation of terms in the sum in 
Eq.(\ref{e:meff}) is possible, and $\langle m_{\beta \beta} \rangle$ could be smaller than
any of the $m_{\nu_i}$.

The nuclear matrix elements, Gamow-Teller and Fermi, appear in the combination
\begin{equation}
 M_{GT}^{0\nu} - \frac{g_V^2}{g_A^2} M_F^{0\nu} \equiv
\langle f | \sum_{lk} H(r_{lk},\bar{E}_m) \tau_l^+  \tau_k^+
\left( \vec{\sigma}_l\cdot\vec{\sigma}_k - \frac{g_V^2}{g_A^2} | i \rangle \right) ~.
\end{equation}
The summation is over all nucleons, $ | i \rangle,  (| f \rangle)$ are the initial 
(final) nuclear states, and $H(r_{lk},\bar{E}_m)$ is the `neutrino potential'
(Fourier transform of the neutrino propagator) that depends (essentially as $1/r$)
on the internucleon distance. When evaluating these matrix elements the short-range
nucleon-nucleon repulsion must be taken into account due to the mild emphasis on
small nucleon separations.

There is a vast literature devoted to the evaluation of these nuclear matrix elements,
going back several decades. It is beyond the scope of the present review to describe
this effort in detail. The interested reader can consult various reviews on the subject,
e.g. \cite{EV,FS98,SC98}. 

Obviously, any uncertainty in the nuclear matrix elements is reflected as
a corresponding uncertainty in the $\langle m_{\beta\beta} \rangle$. There is,
at present, no 
model independent way to estimate the uncertainty, and to check which of the
calculated values are correct. Good agreement with the known $2\nu\beta\beta$
is a necessary but insufficient condition. The usual guess of the uncertainty
is the spread, by a factor of $\sim 3$, of the matrix elements calculated
by different authors. Clearly, more reliable evaluation of the nuclear matrix
element is a matter of considerable importance. (For a recent attempt to reduce
and understand the spread of the calculated values, see \cite{RFSV03}.) 

The $0\nu\beta\beta$ decay is not the only possible observable manifestation of 
lepton number violation. Muon-positron conversion, 
\begin{equation}
\mu^- + (A,Z) \rightarrow e^+ + (A,Z-2) ~,
\label{eq:muon_conv}
\end{equation}
or rare kaon decays $K_{\mu\mu\pi},~K_{ee\pi}$ and $K_{\mu e \pi}$,
\begin{equation}
K^+  \rightarrow  \mu^+ \mu^+ \pi^-~, K^+  \rightarrow  e^+ e^+ \pi^-  ~,
K^+  \rightarrow \mu^+ e^+ \pi^- ~,
\label{eq:kaon}
\end{equation}
are examples of processes that violate total lepton number
conservation and where good limits on the corresponding branching
ratios exist. (See Ref. \cite{Zuber00} for a more complete discussion.)
However, it appears that the   $0\nu\beta\beta$ decay is, at present,
the most sensitive tool for the study of the Majorana nature of neutrinos.

\subsection{Direct measurement of neutrino mass}

Conceptually the simplest way to explore the neutrino mass is to determine its
effect on the momenta and energies of charged particles emitted in weak decays.

In two-body decays, e.g. 
$\pi^+ \rightarrow \mu^+ + \nu_{\mu}$  the analysis is particularly straightforward,
at least in principle. In the system where the decaying pion is at rest, the
energy and momentum conservation requirements mean that
\begin{equation}
m_{\nu}^2 = m_{\pi}^2 + m_{\mu}^2 - 2m_{\pi} \sqrt{ m_{\mu}^2 + p_{\mu}^2}~.
\end{equation}
However, the neutrino mass squared appears as a difference of two very large
numbers. Hence the uncertainties in $m_{\pi}, p_{\mu}$ and even $m_{\mu}$ mean
that the corresponding mass limit is only $m_{\nu} < 170$ keV.

This problem can be avoided by studying the three body decays, in particular
the nuclear beta decay. Near the endpoint of the beta spectrum a massive
neutrino has so little kinetic energy that the effects of its finite mass
become more visible. The electron spectrum of an allowed beta decay is given
by the corresponding phase-space factor
\begin{equation}
\frac{dN}{dE} \sim F(Z,E_e) p_e E_e (E_0 - E_e) [(E_0 - E_e)^2 - m_{\nu}^2]^{1/2} ~,
\label{e:bspect}
\end{equation}
where $E_e, p_e$ is the electron energy and momentum and $F(Z,E_e)$ describes the
Coulomb effect on the outgoing electron. The quantity $E_0$ is the endpoint energy,
the difference of total energies of the initial and final systems.
Clearly, the effect of finite neutrino mass becomes visible if
$(E_0 - E_e) \sim m_{\nu}$, i.e. very near the threshold.

For the case of several massive neutrinos with mixing, the beta decay spectrum is an
incoherent superposition of spectra like (\ref{e:bspect}) with corresponding weights
$|U_{e i}|^2$ for each mass eigenstate $m_{\nu_i}^2$. If the experiment has 
insufficient energy resolution the quantity
\begin{equation}
m_{\nu_e}^{2 {\rm (eff)}} = \sum_i |U_{e i}|^2 m_{\nu_i}^2 
\end{equation}
(using the RPP notation \cite{pdg}) could be determined from the electron
spectrum near its endpoint, 
where the sum is over all the experimentally unresolved
neutrino masses $m_{\nu_i}$.

Based on an upper limit for the $m_{\nu_e}^{2 {\rm (eff)}}$
one can deduce several limits that do not depend on the
mixing parameters $|U_{e i}|^2$. First, at least one of the
neutrinos (i.e. the one with the smallest
mass) has a mass less or equal to that limit, 
$m^2_{\nu_{min}} \leq m^{2{\rm (eff)}}_{\nu_e}$.
Moreover,  if all (with an emphasis on {\it all})  
$|\Delta m_{ij}^2|$ values
are known, an upper limit of {\it all} neutrino masses is
$m_{\nu_{max}}^2 \leq m^{2{\rm (eff)}}_{\nu_e} + \sum_{i < j}
|\Delta m_{ij}^2|$.  
Thus, if we assume that the $|\Delta m_{ij}^2|$ values
deduced from the experiments on solar (and reactor) and atmospheric
oscillation studies
(and include also the LSND result) cover all possibilities, and combine
that knowledge with the  $m_{\nu_e}^{2 {\rm (eff)}}$ limit from
tritium beta decay, we may conclude that no active neutrinos with
mass more than `a few' eV exists.

\section{Experimental Results and Interpretation}

Positive evidence for neutrino mass has so far been obtained only in 
measurements of neutrino oscillations.
As noted above in {\it 2.3}, the neutrino masses
enter only through the differences in squared masses
(i.e., $\Delta m_{ij}^2 = m_i^2-m_j^2$), so although these measurements
provide lower limits to the mass values (e.g., $m_i^2 \ge \Delta m_{ij}^2$)
they do not actually determine the masses. 
As discussed in {\it 2.6} and {\it 2.7}, 
direct neutrino mass measurements and neutrinoless
double beta decay 
allow determination of quantities related to the values of the masses themselves (as
opposed to differences). However, these experiments have so far yielded only limits.
Finally, additional constraints have been derived from measurements
of the cosmic microwave background radiation 
and galaxy distribution surveys through the effect of neutrinos on the
distribution of matter in the early universe. In this section,  we summarize the 
positive observations from oscillation measurements followed by brief discussion of the 
direct measurements, double beta decay searches, and recent constraints from cosmology.

\subsection{Neutrino Oscillation Results}

The first hints that neutrino oscillations actually occur were serendipitously 
obtained through early studies of solar neutrinos and neutrinos 
produced in the atmosphere
by cosmic rays (``atmospheric neutrinos''). In fact, 
the atmospheric neutrino measurements were
a byproduct of the search for proton decay using 
large water \v{C}erenkov detectors. So it is somewhat
ironic that although there was substantial interest 
in searching for neutrino oscillations, the
first evidence for this phenomena came from experiments 
designed for very different purposes.

As shown below,  recent studies definitively establish 
that the solar neutrino flux is reduced due to
flavor oscillations, and so it is now clear
that the first real signal of neutrino oscillations was the long-standing
deficit of solar neutrinos observed by Ray Davis and collaborators 
using the Chlorine radiochemical
experiment in the Homestake mine. While it took almost three decades 
to demonstrate the real origin
of this deficit, the persistent observations by Davis {\it et al.} 
and many other subsequent solar-$\nu$
experiments were actually indications of neutrino oscillations. 
We will discuss this subject in more detail below.

\subsubsection{Atmospheric Neutrinos}

The Kamiokande experiment in Japan \cite{Kam} and the 
IMB experiment in the US \cite{IMB}were
pioneering experimental projects
to develop large volume water \v{C}erenkov detectors 
with the primary goal of detecting nucleon decay,
as predicted by Grand Unified Theories developed in the 1970's \cite{GUT}. 
Although these detectors
were located deep underground to avoid cosmic ray-induced background, they both 
encountered the potential background events produced by atmospheric neutrinos 
(both $\nu_e$ and $\nu_\mu$) that easily
penetrated to these subterranean labs and (rarely) produced energetic events in the huge
detectors. And indeed, both experiments \cite{Kam1,IMB1,Beier} 
(along with the Soudan experiment \cite{Soudan}) observed that the
ratio of $\nu_\mu$-induced events to $\nu_e$-induced events 
was substantially reduced from 
the expected value of $\sim 2$. The decay chain of $\pi^\pm$ produced in the upper 
atmosphere would produce (through the subsequent 
$\mu$-decay) a $\nu_\mu$, $\bar \nu_\mu$,
and a $\nu_e$ (or $\bar \nu_e$). Thus, based on rather simple basic arguments one expects
the ratio of $\nu_\mu$/$\nu_e$ events to be about $\sim 2$ - 
and this is supported by more 
detailed Monte Carlo simulations. The observed values were closer 
to $\sim 1$, which was viewed
as an anomaly for many years. Here again, although 
$\nu$-oscillations could clearly cause this
anomaly there was not enough corroborative evidence to substantiate this explanation.

\begin{figure}
\leftline{
\includegraphics[angle=-90,width=5.3 in]{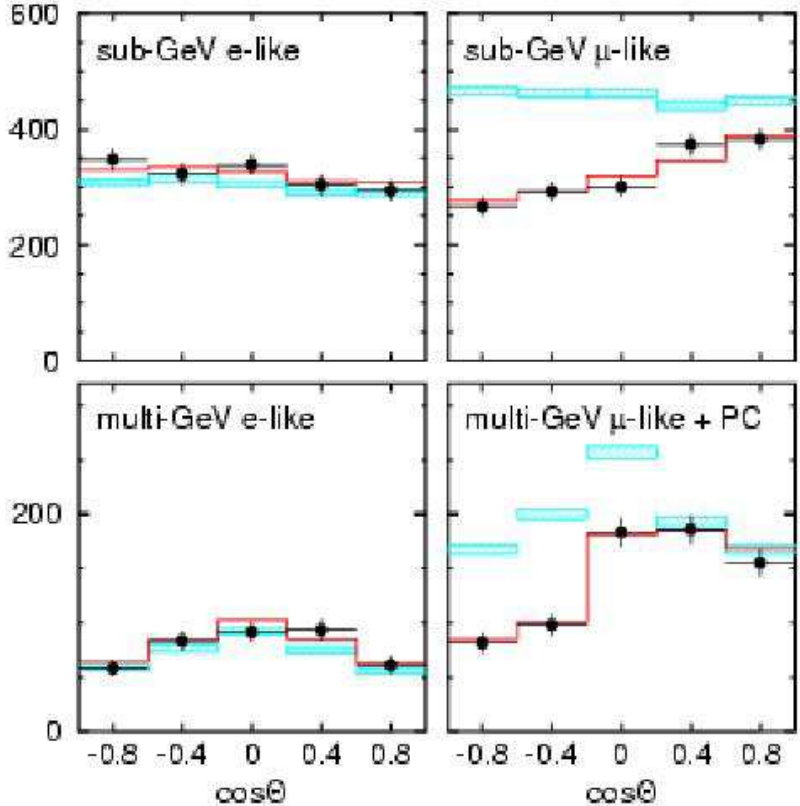}}
\vspace{0.5 in}
\caption[]{Distribution of observed atmospheric neutrino events 
vs. zenith angle from the 
SuperKamiokande experiment, compared with Monte Carlo simulations \cite{Kate}. 
The blue hatched region
represents the prediction without neutrino oscillations 
and the red line includes the effect
neutrino oscillations. (PC means `partially contained' events.)
}
\label{fig:SK_atmos}
\end{figure}

However, the situation dramatically changed in 1998 
when the larger experiment, Super-Kamiokande,
reported a clearly anomalous zenith angle dependence of the  
$\nu_\mu$ events \cite{SK_atmos}. The
measurements indicated
a deficit of upward-going $\nu_\mu$-induced events 
(produced $\sim 10^4$~km away at the opposite
side of the earth) relative to the downward-going events (produced $\sim 20$~km above). 
The $\nu_e$ events displayed a normal zenith angle behavior consistent with Monte Carlo 
simulations. Since at that time (1998) the solar-$\nu$ problem was
still unresolved, these data represent the first 
really solid evidence for $\nu$-oscillations.
More recent measurements \cite{Kate} are displayed in 
Fig.~\ref{fig:SK_atmos}, and the conclusion that flavor oscillations are 
responsible is essentially inescapable. 
Moreover, the deduced values of $\sin^2 2\theta > 0.90$ (90\% C.L.) 
indicate a surprisingly strong
mixing scenario where the muon-type neutrino seems 
to be a fully-mixed superposition of all three mass 
eigenstates. (This situation is completely contrary 
to the quark sector, where the mixing between
generations is generally small.) The most recently reported  
value of $\Delta m^2$ derived from the
Super-Kamiokande results is 0.0020 eV$^2$ \cite{SK03}
(for the error bars, see Table 1).

The observation of the angular distribution of upward-going muons produced
by atmospheric neutrinos in the rock below the MACRO detector \cite{Ambrosio01}
supports the conclusion that the observed effect is due
to the $\nu_{\mu} \rightarrow \nu_{\tau}$ oscillations, and disfavors
$\nu_{\mu} \rightarrow \nu_{sterile}$ assignment.

Although the parallel effort to detect neutrino disappearance 
at nuclear reactors had made steady
progress (setting upper limits and establishing exclusion plots) for many years, these
experiments made a strong contribution at this point: 
the failure to observe $\bar \nu_e$ disappearance
at CHOOZ \cite{CHOOZ} and Palo Verde \cite{PV} in the region near
 $\Delta m^2 \simeq 0.0020~$eV$^2$ implies that the $\nu_\mu$ disappearance observed by 
Super-Kamiokande does {\it not} involve substantial $\nu_e$ appearance. (This inference
assumes that $m_\nu=m_{\bar \nu}$ for each eigenstate as required by $CPT$ invariance.)
Thus, it would seem that the $\nu_\mu$'s must
be oscillating into $\nu_\tau$ or other more exotic particles, 
such as ``sterile'' neutrinos.

\subsubsection{Solar Neutrinos}

\begin{figure}[t]
\includegraphics[angle=-90,width=5in]{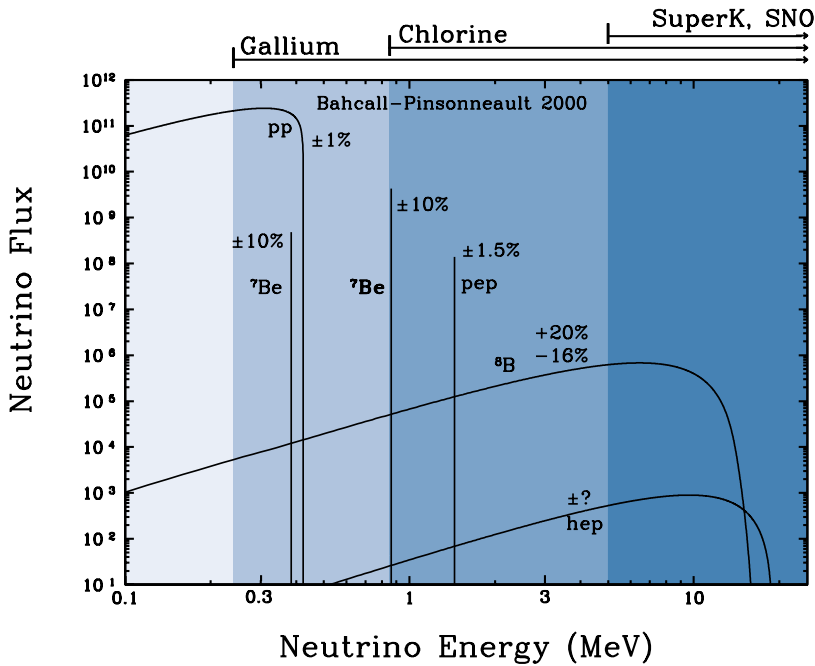}
\caption[]{Energy distribution of the flux of solar neutrinos 
predicted by the standard solar
model \cite{bp00}, as computed by J. Bahcall \cite{solarspec}. The ranges of energies
associated with the various experiments
are indicated at the top of the figure.
}
\label{fig:solarspec}
\end{figure}

The interpretation of solar neutrino measurements
involves substantial input from solar physics and the nuclear physics involved in
the complex chain of reactions that together are termed the 
``Standard Solar Model'' (SSM) \cite{bp00} .
The predicted flux of solar neutrinos from the 
SSM is shown in Fig.~\ref{fig:solarspec} as a 
function of neutrino energy. 
The low energy $p$-$p$ neutrinos are the most abundant and,
since they arise from reactions that are responsible 
for most of the energy output of the sun,
the predicted flux of these neutrinos is constrained 
very precisely ($\pm 2\%$) by the solar luminosity.
The higher energy neutrinos are more accessible experimentally, but the fluxes are less
certain due to uncertainties in the nuclear and solar physics required to compute them.

The early measurements included
radiochemical experiments sensitive to integrated $\nu_e$ flux such as the
 Chlorine \cite{Davis} (threshold energy 0.814~Mev) and Gallium \cite{Gallium}
(0.233~MeV) experiments. Live counting was developed by the Kamiokande 
\cite{Kam_solar} and then 
the SuperKamiokande \cite{SK_solar} experiments,
based on neutrino-electron scattering, 
enabling measurements of both the flux and energy spectrum.
Together these experiments sampled the
solar neutrino flux over a wide range of energies. 
As can be seen in Fig.~\ref{fig:allsolar} below, all these
experiments reported a substantial deficit in neutrino flux relative to the SSM.
While it was realized that neutrino oscillations could be 
an attractive solution to this problem, it was
problematic to establish this explanation with certainty 
due to the dependence on the SSM,
its assumptions, and sensitivity to input from nuclear and solar physics.

However, it is now clear that the precise measurements of solar neutrino fluxes
over a range of energies coupled with the amazing flavor 
transformation properties of neutrinos 
validates the SSM in a beautiful and satisfying manner.
The solar-$\nu$ experiments and the SSM have been reviewed in detail
elsewhere \cite{pdg}, and so it is not appropriate to repeat the details here.
We present these solar neutrino flux
measurements together in Fig.~\ref{fig:allsolar}, where we also display the 
excellent agreement with the best fit solution 
to the combined solar-$\nu$ and KamLAND reactor data \cite{SNO_salt}
to demonstrate the remarkably coherent 
picture that is strongly supported by these impressive experimental measurements and the
associated SSM input.

\subsubsection{SNO and KamLAND}

With the advent of the new millenium, 
the stage was set for a synthesis of the
study of solar neutrinos using a powerful new
water \v{C}erenkov detector (Sudbury Neutrino Observatory, SNO)
with the study of the disappearance of
reactor antineutrinos using a large scintillation detector located deep underground
(KamLAND). The results of these experiments provide definitive evidence that the
solar deficit is indeed due to flavor oscillations, and that this effect is demonstrable
in a ``laboratory'' experiment on earth.

The SNO experiment combines the now high-developed capability 
of water \v{C}erenkov detectors with
the unique opportunities afforded by using deuterium to detect the solar neutrinos 
\cite{SNO1,SNO2}.
Low energy
neutrinos can dissociate deuterium via the charged current (CC) reaction
\begin{equation}
\nu_e + d \rightarrow e^- + 2p
\end{equation}
or the neutral current (NC) reaction 
\begin{equation}
\nu_\ell + d \rightarrow \nu_\ell + p + n \> .
\end{equation}
Only $\nu_e$ can produce the CC reaction, 
but all flavors $\ell= e, \mu, \tau$ can contribute
to the NC rate. The CC reaction is detected via 
the energetic spectrum of $e^-$ which closely
follows the ${}^8$B solar $\nu_e$ spectrum. 
The NC reaction involves three methods for detection
of the produced neutron: (a) capture on deuterium and detection 
of the 6.25 MeV $\gamma$-ray,
(b) capture on Cl (due to salt added to the D${}_2$O) 
and detection of the 8.6 MeV $\gamma$-ray,
or (c) capture in ${}^3$He proportional counters immersed in the detector. 
There are also some events associated with the elastic scattering 
of the solar-$\nu$ on $e^-$ in the detector
which is dominated by the charged current reaction (again only $\nu_e$) but has
some $\sim 20\%$ contribution from neutral currents (all flavors equally contribute).

\begin{figure}[h]
\centerline{
\includegraphics[width=4.5 in]{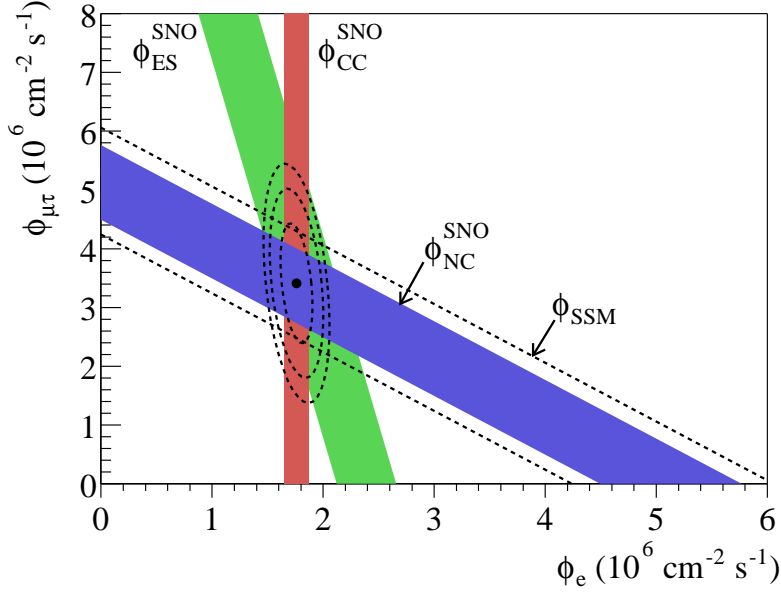}}
\caption[]{Measured solar neutrino fluxes from SNO 
for the NC and CC processes \cite{SNO2},
along with elastic scattering events (ES) and the SSM prediction \cite{bp00}.
}
\label{fig:SNO}
\end{figure}

The SNO collaboration has published data on the CC and NC rate 
(from processes (a) and (b)). Additional data
from the NC process (c) will be forthcoming in the future. 
Nevertheless, the reported results (see Fig.~\ref{fig:SNO}) demonstrate very clearly
that the total neutrino flux ($\nu_e + \nu_\mu + \nu_\tau$ as
determined from NC) is in good agreement with the SSM, but that the $\nu_e$ flux is 
suppressed (as determined from CC). This represents rather definitive evidence that
the $\nu_e$ suppression is due to flavor-changing processes that convert the $\nu_e$
to the other flavors, as expected from $\nu$-oscillations. Furthermore, the observed
value of $\nu_e$ flux and the observed energy spectrum, when combined with the other
solar-$\nu$ measurements strongly favor another large mixing angle scenario at a lower
value of $\Delta m^2 \sim 10^{-5}$ eV$^2$.

\begin{figure}[b]
\centerline{
\includegraphics[width=4.5 in]{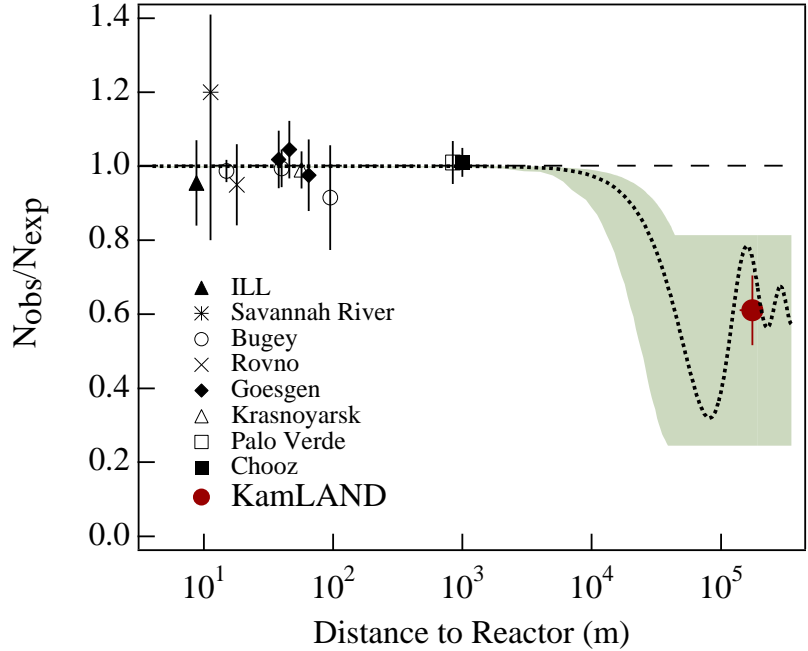}}
\caption[]{Ratio of observed to expected rates 
(without neutrino oscillations) for reactor neutrino
experiments as a function of distance, including 
the recent result from the KamLAND experiment
\cite{KamLAND}.
The shaded region is that expected due to neutrino oscillations with large mixing angle 
parameters as determined from solar neutrino data.
}
\label{fig:KamLAND}
\end{figure}

\begin{figure}[h]
\centerline{
\includegraphics[angle=-90,width=3.7 in]{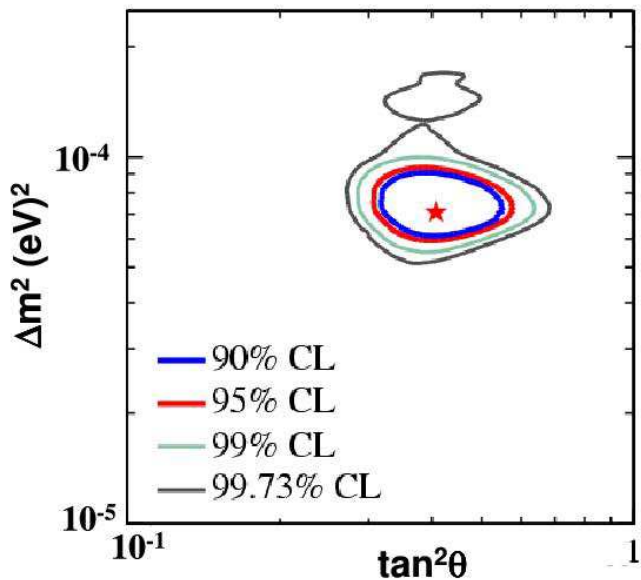}}
\vspace{0.2 in}
\caption[]{Region of parameter space constrained by simultaneous fit to solar-$\nu$
and KamLAND data, from \cite{SNO_salt}. The best fit values are $\Delta m^2 = 7.1 \times
10^{-5}$~eV${}^2$ and $\tan^2 \theta = 0.41$.
}
\label{fig:Bahcall_fit}
\end{figure}

The KamLAND experiment \cite{KamLAND} represents 
a major advance in the development of reactor 
$\bar \nu_e$ measurements. In order to reach the low values of 
$\Delta m^2 \sim 10^{-5}$ eV$^2$ indicated by the solar-$\nu$ data, distances
of order $\sim 200$~km are necessary (given the fixed range of $\bar \nu_e$ energies from
reactors). The loss of rate due to $1/r^2$ scaling is severe, which requires substantial
increases in source strength (i.e., reactor power) and detector size. 
Such a huge detector,
sensitive to low-energy inverse beta-decay events from reactor $\bar \nu_e$, would 
be very susceptible to background from cosmic radiation and so must
be located deep underground. By fortunate coincidence, the old Kamiokande site 
in Japan is very
deep ($\sim 1000$ mwe) and located an average distance of $\sim 200$~km 
from a substantial number of large
nuclear power reactors. Thus the KamLAND experiment, 
a large liquid scintillator detector,
was built at this site to study the disappearance of $\bar \nu_e$ from nuclear reactors.
For the first time in the long history of reactor $\bar \nu_e$ experiments (dating
back to the original discovery of the 
$\bar \nu_e$ by Reines and Cowan \cite{Reines}) a substantial
deficit in event rate was observed (Fig.~\ref{fig:KamLAND}). 
In fact this deficit is just as predicted by the solar
solution to the solar-$\nu$ oscillation solution, 
and the more precisely constrained values of
$\Delta m^2$ and $\sin^2 2 \theta$ from a global analysis 
of the solar-$\nu$ and KamLAND data
are shown in Fig.~\ref{fig:Bahcall_fit}. 

\begin{figure}[h!]
\vspace{0.15 in}
\centerline{
\includegraphics[angle=90,width=4. in]{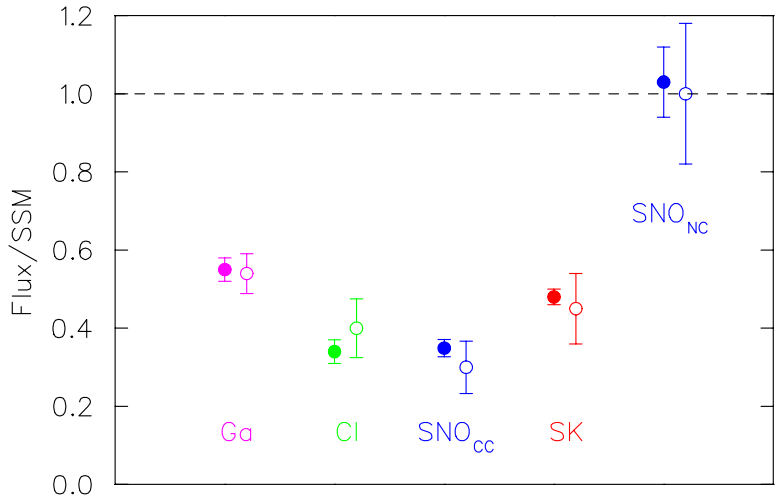}}
\caption[]{Ratio of solar neutrino flux to SSM (without neutrino oscillations) for 
various experiments (see  text). Filled circles are 
experimental data (with experimental uncertainties only)
and open circles are theoretical expectations based on SSM with best fit parameters
to KamLAND and solar-$\nu$ data (uncertainties from SSM and oscillation fit combined).
All charged current experiments show a substantial deficit 
and all are in excellent agreement
with the expected values.
}
\label{fig:allsolar}
\end{figure}

As mentioned previously, Fig.~\ref{fig:allsolar} shows 
a summary of the solar-$\nu$ data compared with the SSM with and without neutrino 
oscillations. The plotted experiments are: Ga, 
combined Gallium measurements from GALLEX and SAGE
\cite{Gallium}; 
Cl, Chlorine measurement from Homestake mine \cite{Davis}; 
SNO, Sudbury Neutrino Observatory 
(CC and NC)\cite{SNO_salt,SNO1,SNO2};
and SK, Super-Kamiokande\cite{SK_solar}.
In this plot one can clearly see the decreasing survival fraction with increasing
energy in the progression Ga$\rightarrow$Cl$\rightarrow$SNO$_{\rm CC}$
(all sensitive only to the $\nu_e$ component). 
The SNO$_{\rm NC}$ measurement
shows no suppression, whereas the SK data exhibit the intermediate suppression due to the
partial contribution of NC events to their elastic scattering signal.

\begin{figure}[h!]
\vspace{0.25 in}

\centerline{
\includegraphics[angle=90,width=4. in]{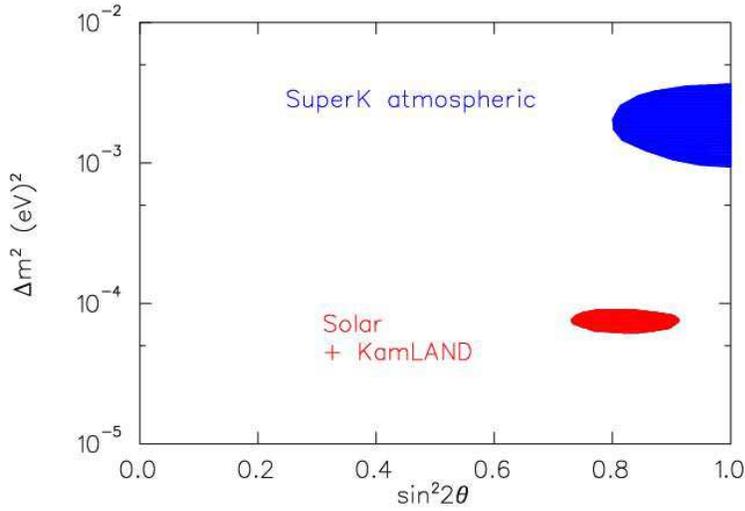}}
\caption[]{Allowed regions of parameter space (90\% C.L.)
determined by atmospheric neutrino 
measurements \cite{SK03}
($2 \leftrightarrow 3$ mixing) and by solar-$\nu$ 
and reactor-$\bar \nu$ measurements \cite{SNO_salt}
($1 \leftrightarrow 2$ mixing).
}
\label{fig:atmos_solar}
\end{figure}

In summary, the experimental studies of solar neutrinos, 
atmospheric neutrinos, and reactor 
antineutrinos have established neutrino oscillations with two different mass scales,
$\Delta m^2 \sim 2.0 \times 10^{-3}$~eV$^2$ and 
$\Delta m^2 \sim 7.1 \times 10^{-5}$~eV$^2$, both
with large associated mixing angles. 
The allowed regions are shown in Fig.~\ref{fig:atmos_solar}, and together 
these experiments constrain many of the matrix elements in the $3 \times 3$ mixing
matrix for the neutrinos along with the mass differences 
$\Delta m_{12}^2$ and $\Delta m_{32}^2$.
The results for these parameters are listed in Table~\ref{tab:params}.


\vspace{0.25 in}

\begin{table}[h!]
\caption{Neutrino Oscillation Parameters Determined From Various Experiments (2003)}
\label{tab:params}
\begin{center}
\begin{tabular}{ccll}
\hline
Parameter & Value $\pm 1 \sigma$ & Reference & Comment\\

\hline \hline
$\Delta m_{12}^2$ & $7.1^{+1.2}_{-0.6} \times 10^{-5}$~eV$^2$& \cite{SNO_salt}&\\
 $ \theta_{12} $ & ${32.5^{\circ +2.4}_{-2.3}}$ & \cite{SNO_salt} & For $\theta_{13}=0$\\
 $\Delta m_{32}^2$  & $2.0^{+0.6}_{-0.4} \times  10^{-3}$~eV$^2$ & \cite{SK03}& \\
  $\sin^2 2 \theta_{23} $  & $>0.94$ & \cite{SK03} &For $\theta_{13}=0$\\
$\sin^2 2\theta_{13}$  & $<0.11$ & \cite{CHOOZ} & For $\Delta m^2_{atm} =
2\times 10^{-3}$ eV$^2$ \\

\hline \hline

\end{tabular}
\end{center}
\end{table}

\newpage

\subsubsection{LSND}

\begin{figure}
\includegraphics[angle=-90,width=4.8 in]{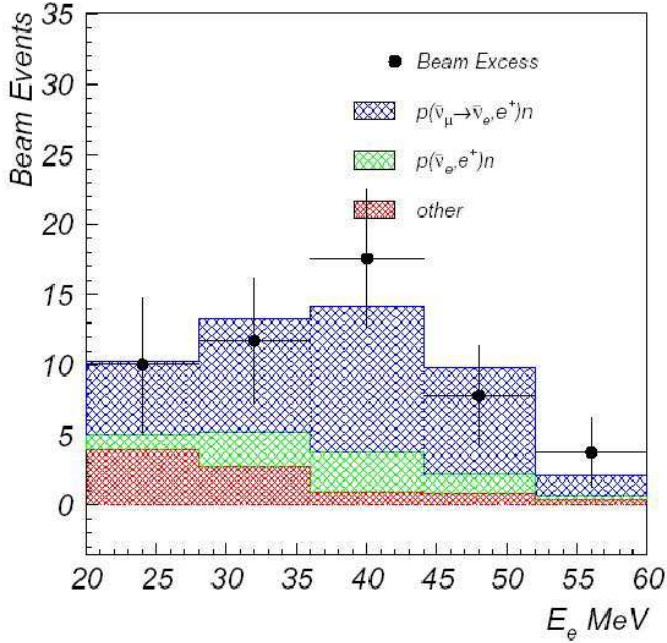}
\vspace{0.5 in}
\caption[]{Observed spectrum of positron energy 
from beam-induced events in the LSND experiment
\cite{LSND}. The black dots are the experimental data, 
the lower histogram is the estimate
of various backgrounds, the middle histogram includes the estimated $\bar \nu_e$
contamination from the source, and the top histogram includes the neutrino oscillation
hypothesis.
}
\label{fig:lsnd}
\end{figure}

There is one other experiment that claims to observe neutrino 
oscillations: the Liquid
Scintillator Neutrino Detector (LSND) \cite{LSND} at Los Alamos. 
In this experiment, the neutrino source was
the beam dump of an intense 800 MeV proton beam where 
a large number of charged pions were created
and stopped. Since the $\pi^-$ capture on nuclei with very high probability, 
essentially only the $\pi^+$ decay,
producing $\nu_\mu$ and then $\bar \nu_\mu$ and $\nu_e$ 
(from $\mu^+$ decay). These neutrinos all
have very well defined energy spectra 
(from decays of particles at rest) and note that there
are no $\bar \nu_e$ produced in this process.
The 160 ton detector is then
used to search for $\bar \nu_e$ events via inverse 
beta decay on protons at a distance of 30~m from the
neutrino source. The experiment detected an excess of $87.9 \pm 22.4 \pm 6.0$
events corresponding to an oscillation probability of $0.264 \pm 0.067 \pm 0.45$\%.
(Note that such an appearance experiment affords access to very small mixing parameters.)
The observed spectrum of events is shown in Fig.~\ref{fig:lsnd}.
Other experiments, especially the KARMEN accelerator experiment 
\cite{Karmen} and the Bugey
reactor experiment \cite{Bugey}, rule out much of the
allowed region of parameter space but there 
is a small region remaining  at 90\% confidence in the
mass range $0.2 < \Delta m^2 < 10$~eV$^2$, indicating a minimum mass of $m_\nu>0.4$~eV 
(see Fig.~\ref{fig:miniboone}). It is also significant that the KARMEN experiment studied a
shorter baseline, which seems to rule out 
the possibility that the $\bar \nu_e$ are produced
at the source.

Given the well established values of $\Delta m^2$ in Table~1, it is not possible to
form a third value of $\Delta m^2$ consistent with the LSND data (note that 
$\Delta m_{12}^2 + \Delta m_{23}^2 + \Delta m_{31}^2 =0$). 
So one would need to either break
CPT invariance (allowing $m_\nu \neq m_{\bar \nu}$), or invoke additional ``sterile'' 
neutrinos that do not have the normal weak interactions. 
In addition, the LSND range of $\Delta m^2$ is marginally at variance with recent
studies of the cosmic microwave background (see below).
Therefore, it is of great 
importance to attempt to obtain independent verification 
of this result (see {\it 4.1} below).

\subsection{Direct Mass Measurements}

Beta decays with low endpoints, in particular tritium $(Q = 18.6$ keV), have
been used  for a long time in attempts to measure or constrain
neutrino mass. ($^{187}$Re with $Q = 2.5$ keV
has been also explored recently \cite{Re187}, but the sensitivity is still 
only 21.7 eV, a factor of about ten worse than for tritium.)

There are several difficulties one has to overcome to reach sensitivities to
small neutrino masses in the study of the electron spectrum
of $\beta$ decay. Obviously, the mass sensitivity can be only as good
as the energy resolution of the electron spectrometer. Also, since by
definition the energy spectrum vanishes at the endpoint, the fraction of
events in the interval $\Delta E$ near the $Q=E_0 - m_e$ kinetic energy
endpoint decreases rapidly, as $(\Delta E/Q)^3$. Finally, since the decaying
system is a molecule or at best an atom, one has to take into account 
the possibility that the sudden change of the nuclear charge causes excitations
or ionization of the electron cloud, i.e., the presence of multiple endpoints.

\begin{figure}
\includegraphics[angle=90,width=4.5 in]{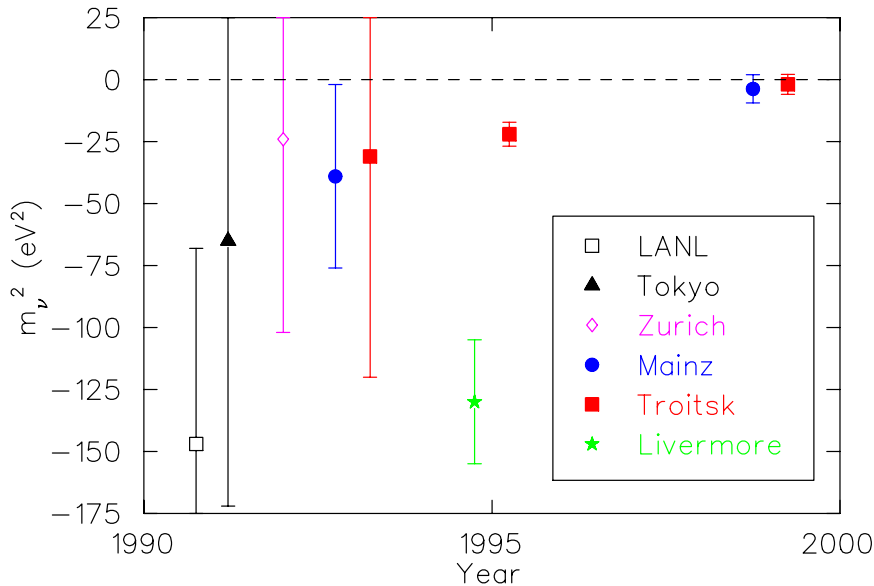}
\caption[]{Results of measurements of $m_{\bar \nu_e}^{2 {\rm (eff)}}$ 
from tritium $\beta$-decay experiments since 1990 \cite{pdg}, showing
the steady improvement in the achieved precision.
}
\label{fig:direct}
\end{figure}

Although there are as yet no positive results from direct 
neutrino mass measurements, these 
efforts have made (and continue to make) substantial progress 
(see Fig.~\ref{fig:direct}). 
Curiously, many past experiments have been apparently plagued by systematic effects
that tended to mimic a negative $m_{\bar \nu_e}^{{\rm (eff)} \ 2}$.
As the experiments were improved, it seems that these problems have been largely
overcome.
Two recent 
tritium $\beta$ decay experiments have
reached impressive results \cite{tritium1,tritium2}
quoting 95\% upper limits for the effective mass parameter 
$m_{\bar \nu_e}^{{\rm (eff)}} = \sqrt{\sum_i |U_{ei}|^2 m_i^2}$
of 2.5~eV and 2.8~eV. However, these values are now 
at the level where atomic and chemical
effects on the phase space distributions are significant, and
it seems that some of these experiments \cite{tritium2}
still occasionally observe anomalous structures near the 
endpoint. This will likely be a 
substantial challenge
for future measurements of this type.

Constraints from direct mass measurements
also exist on $m_{\nu_\mu}$ and $m_{\nu_\tau}$, but these are much larger
(see Section {\it 2.7} for explanation). Given reasonable
assumptions, it seems that the oscillation 
results discussed above and the constraints from
tritium decay would imply that the mass values 
are far below the direct measurements for 
$\nu_\mu$ and $\nu_\tau$ so we do not dicuss them in detail here.

\subsection{Double Beta Decay}

Over the last decade, the methodology for double beta decay 
experiments has markedly improved. 
Larger volumes of high-purity enriched materials 
are being utilized, and careful selection
of materials along with deep-underground siting have lowered backgrounds and increased 
sensitivity. The most sensitive experiments use the isotopes
${}^{76}$Ge,  ${}^{100}$Mo, $^{116}$Cd, ${}^{130}$Te, and ${}^{136}$Xe. 
For ${}^{76}$Ge, the lifetime limit has reached an
impressive value exceeding $10^{25}$ years. 
(The experimental results are listed in \cite{pdg},
for the latest review of the field, see \cite{EV}.)
The conversion of the observed lifetime limits 
to effective neutrino mass values 
requires use of calculated nuclear matrix elements, and there is some uncertainty 
associated with them. Nevertheless, the experimental lifetime limits have been
interpreted to yield effective mass limits of typically a few eV
and in $^{76}$Ge of 0.3-1.0 eV. One recent
report \cite{Klapdor}, analyzing the ${}^{76}$Ge 
data from the Moscow-Heidelberg experiment,
claims to observe a positive
signal corresponding to the effective mass $\langle m_{\beta\beta} \rangle 
= 0.39^{+0.17}_{-0.28}$~eV. 
That report has been followed by a lively discussion
\cite{Aalseth,Feruglio,Klapdor2,Bakalyarov}. If this finding were to be confirmed 
then it would be a major advance in our knowledge of neutrino properties,
and in particular it would not only prove that neutrinos are Majorana particles,
but it would also strongly indicate that neutrinos follow a degenerate mass pattern,
where $\Delta m^2 / m^2 \ll 1$.

\subsection{Cosmological Constraints}

In the early universe, when the temperature was $T>1$ MeV, 
the high density of particles
allowed weak interactions to occur prolifically leading to 
a substantial density of neutrinos.
As the universe cooled to $T<1$~MeV, these reactions 
became much slower than the expansion
rate and the neutrinos decoupled from the remaining 
ionized plasma and radiation (photons).
Much later ($\sim 100,000$ years), the universe cooled 
enough that atoms formed and the radiation
decoupled from the matter. The cosmic microwave background (CMB) 
is the further cooled (through
expansion) relic of this period, and contains information 
on the distribution of matter at that
time in the history of the universe. The presently observed distribution
of matter (through high resolution galaxy surveys) 
and distribution of radiation (CMB) would
both be affected by
the presence of massive neutrinos in the early universe. Although the power spectra
of the CMB and the density fluctuations are both sensitive to massive neutrinos, a
combined analysis of both observables is especially effective in addressing the existence
of massive neutrinos \cite{BGGM,Hu}.
Thus comparison of the power spectrum of CMB with the observed distribution
of galaxies can provide information on the sum (over all flavors) 
of light neutrino masses.
An analysis of the recent WMAP data \cite{WMAP} yields the result
$\sum_f m_{\nu_f} < 0.7$~eV (95\% confidence). 
It is interesting to note that this significantly constrains the remaining region of parameter 
space allowed by LSND. (In addition, interpretation of the LSND result in light
of cosmological constraints requires careful
consideration of issues related to 
the behavior (e.g. thermalization) of sterile neutrinos in the early universe. \cite{beacom})
However, it has
been argued that the obtained limit also relies 
upon input from Lyman-$\alpha$ forest measurements,
and that a somewhat less restrictive limit should be quoted \cite{Elgaroy,Hannestead1}
Other recent work \cite{SDSS03} also obtains a less restrictive limit without the use
of strong prior on galaxy bias.

\section{Near-term Future}

The recent discoveries and revolutionary breakthroughs 
in the study of neutrino properties
have motivated a new generation of experimental efforts aimed at resolving the remaining
issues and establishing new launching points for future explorations. These near-term
plans and proposals are, for the most part, 
initiatives that advance the themes we have emphasized above.
Neutrino oscillation studies are planned to
address the LSND result, higher precision measurements 
of $\theta_{23}$ and $\Delta m_{23}^2$,
and determination of $\theta_{13}$. Direct mass measurements are aimed
at reducing (or discovering) the absolute value of neutrino masses and establishing the
hierarchical nature of the neutrino mass spectrum. And future experiments in double beta
decay hope to demonstrate the Majorana nature of neutrinos and constrain (or discover)
the mass scale. Finally, the ever increasing precision of cosmological constraints from
measurements of the cosmic micowave background promise to provide tighter limits
on neutrino masses.

\subsection{MiniBooNE}

\begin{figure}
\centerline{
\includegraphics[width=3.5 in]{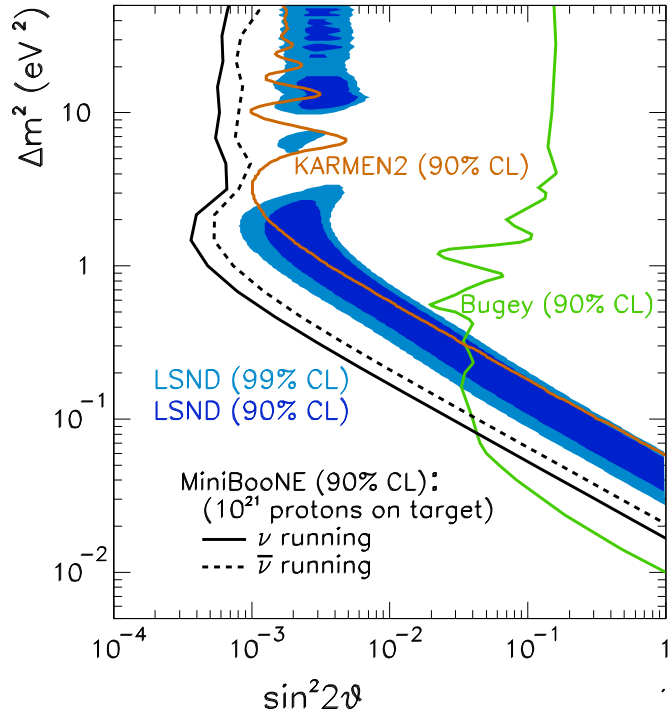}}
\caption[]{Projected sensitivity \cite{miniboone} of 
the MiniBooNE experiment compared with 
LSND and other previous experiments.
}
\label{fig:miniboone}
\end{figure}

A major near-term priority for the field is to resolve 
the issue of the validity of the results
of the LSND experiment. To this end, the Mini-Boone experiment \cite{miniboone}
has been built at FermiLab with 
the main goal to explore the same region of parameter 
space with higher sensitivity. This experiment
uses a new 0.5-1.5 GeV high-intensity neutrino source 
and a  800 ton mineral oil-based \v{C}erenkov detector
located at 500~m from the source to search 
for the oscillation modes $\nu_\mu \rightarrow \nu_e$
and at a later stage $\bar \nu_\mu \rightarrow \bar \nu_e$. 
The projected sensitivity for 2 years running at full
beam intensity in each mode is shown in Fig.~\ref{fig:miniboone}. 
This experiment has the potential to test the
validity of the LSND results with high sensitivity 
and to explore the possible role of $CPT$ violation
by studying both $\nu_e$ and $\bar \nu_e$ appearance. 
If an effect is seen the collaboration plans
to mount another detector at further distance for additional studies.
The initial run with the $\nu_{\mu}$ beam began in 
the Fall of 2002 and first results on
oscillations are expected in 2005.

\subsection{Determination of $\Delta m_{32}^2$ and $\theta_{23}$}

The phenomenon of neutrino oscillation
observed by Super-Kamiokande in the atmospheric neutrino experiment 
requires further study to
precisely determine the mixing parameters $\Delta m_{32}^2$ and $\theta_{23}$. 
There are two long baseline accelerator
experiments with prospects for results in the near term: K2K in Japan and MINOS in 
the US.

\begin{figure}
\includegraphics[angle=-90,width=4. in]{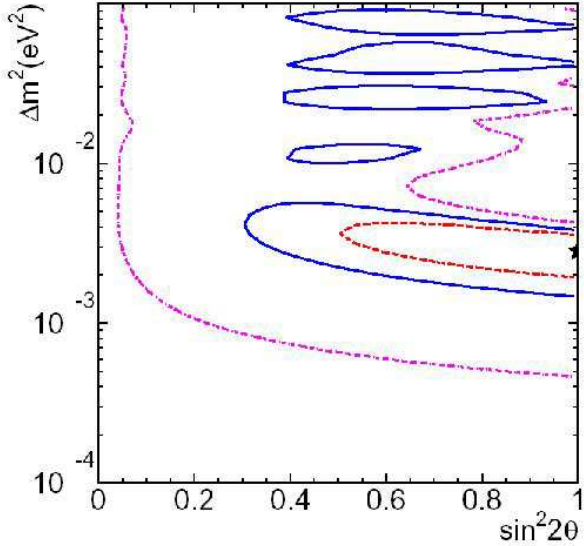}
\vspace{0.4 in}
\caption[]{Allowed regions from analysis of the first results 
from the K2K \cite{K2K} experiment,
representing about 1/2 of the expected total future luminosity. Dashed,
solid and dot-dashed lines are 68.4\%, 90\% and 99\% C.L. contours,
respectively. The best fit point is indicated by the star.}
\label{fig:k2k}
\end{figure}

The K2K experiment utilizes the 12~GeV proton beam at 
KEK to produce a $\nu_\mu$ beam with average
energy 1.3~GeV which is directed at the Super-Kamiokande 
detector a distance of 250~km \cite{K2K}. Use of GPS
receivers enables clean selection of events with the proper 
time relative to the beam spill.
Careful monitoring of the muons in the $\pi^+$-decay 
region ensures proper aiming of the beam over the long
baseline to the detector. In addition, a combination 
of detectors located 300~m downstream of the
$\pi^+$ production target is used to measure 
the flux and energy spectrum of the $\nu_\mu$ beam.
During the first data run in 1999-2001, 
the experiment detected 56 fully-contained $\nu_\mu$ 
events, compared to the expected $80.1^{+6.2}_{-5.4}$ events 
without neutrino oscillations. Thus,
so far the K2K experiment appears to confirm 
the oscillation interpetation of the observed
anomaly in the atmospheric neutrino distribution with high probability. 
By combining the observed energy spectrum and the reduced event rate, the allowed
regions shown in Fig.~\ref{fig:k2k} are obtained. The experimental plan
is to continue running and roughly double this data sample.

The MINOS experiment uses the Main Injector proton beam 
at Fermilab to produce a $\nu_\mu$ beam
in the energy range 3-18~GeV directed at a large detector located in the Soudan mine at a
distance of 735~km \cite{MINOS}. The lowest energy neutrino 
beam will be optimal for studying the region
of $\Delta m_{32}^2 \simeq 2.0 \times 10^{-3}$~eV$^2$. 
This experiment also uses a near detector
(1~kton) in addition to the far detector (5.4~kton). 
The large far detector consists of 486 layers
of alternating steel plates with finely (4~cm) 
segmented plastic scintillator planes. The steel
is magnetized by the return flux of a coil located 
on the detector axis, which enables measurement
of muon momentum. With 10~kton-years of exposure 
at full luminosity, the MINOS experiment should 
easily confirm the $\nu_\mu$ disappearance observed 
with atmospheric neutrinos and determine the
mass parameter $\Delta m_{32}^2$ to about 10\%. 
By searching for $\nu_e$ appearance the MINOS experiment
will be sensitive to $\theta_{13}$ in the region $\sin^2 \theta_{13} > 0.15$ over the 
presently allowed range of $\Delta m_{32}^2$, a factor of $\sim 2$ lower than
the CHOOZ limit. 
The far detector is complete and operating.
It is expected that the MINOS experiment will receive the neutrino beam and start
operation in 2005.

There is also a long baseline facility from CERN 
to the Gran Sasso Laboratory ($L \simeq 730$~km).
This is a higher energy beam ($\bar E_\nu \simeq 17$~GeV), 
suitable for production of $\tau$'s associated
with $\nu_\mu \rightarrow \nu_\tau$ oscillations. 
There will be two detectors at Gran Sasso for this
study: OPERA and ICARUS \cite{TAUP}. The OPERA experiment 
will employ photographic emulsion to identify
the `kinks' associated with the short (few $\mu$m) $\tau$ decays. 
The ICARUS experiment utilizes
a liquid argon time projection chamber to kinematically 
reconstruct the $\tau$ leptons. Both
experiments expect 1-5 events per year to firmly establish 
this oscillation mode, and will begin operation in 2006.
Future experiments providing  precision measurements of the 
corresponding parameters are highly desirable.

\subsection{Studies of $\theta_{13}$, Neutrino Mass Hierarchy, and $CP$ Violation}

The role of mixing between the first and third generation 
neutrino mass eigenstates is pivotal 
in terms of the phenomenological consequences. 
The possibility of $CP$ violation and implications
for leptogenesis scenarios in generating the matter-antimatter 
asymmetry in the universe require
mixing between these two states. Therefore, 
establishment of non-vanishing mixing (i.e., 
$\theta_{13} \neq 0$) is of paramount importance 
and substantial experimental efforts are 
planned to address this issue.

\subsubsection{High-precision Reactor Neutrino Experiments}

When completed, the KamLAND experiment will significantly 
reduce the allowed region for $\Delta m_{12}^2$
and $\tan^2 \theta_{12}$ relative to the present 
results shown in Fig.~\ref{fig:Bahcall_fit}. 
The next major goal
for the reactor neutrino program will be to attempt 
a measurement of $\sin^2 \theta_{13}$. 
As can be seen from Eq.~(\ref{e:oscs3}) and as discussed
in section {\it 2} above, such experiments have the potential to determine
$\theta_{13}$ without ambiguity from $CP$ violation or matter effects.
The strategy is to capitalize on the success of CHOOZ and KamLAND 
to build a new experiment
at the $1.5-2$~km distance now indicated 
by the Super-Kamiokande atmospheric neutrino results.
Obtaining the necessary statistical precision 
on this challenging disappearance experiment
will require large reactor power ($> 5$~GWth) and large detector size 
(probably $\sim 100$~ton). Systematic
errors must be carefully controlled through reduced
background and utilization of a two detector scheme (probably in a 
configuration with a near detector at a close $\sim 100-200$~m distance from the
reactor(s) in addition to the far detector at $\sim 2$~km). 
Backgrounds must be controlled through sufficient depth underground ($>300$~m to reduce 
cosmic ray induced spallation products), careful choice of detector materials, and 
passive shielding (e.g., buffer of water or mineral oil) 
to reduce events due to radioactive
decays in the surrounding rock and other material.
Comparison of the rates and 
observed spectral shapes in the two detectors 
will reduce the sensitivity to reactor source
uncertainties and provide substantial improvements 
in the sensitivity to reactor antineutrino
disappearance in this region.
With the anticipated statistical and systematic 
uncertainties held to a total of order 1\%, the
sensitivity will reach about $\sin^2 2\theta_{13} \simeq 0.01-0.02$ at the 
optimal value of $\Delta m^2 \sim 2.0 \times 10^{-3}$~eV$^2$ 
derived from the Super-Kamiokande
results. Significant efforts are presently underway 
to optimize the experimental design and 
select appropriate sites for this future experiment. \cite{Huber,Suekane,Goodman}

\begin{figure}[b]
\includegraphics[angle=-90,width=5.2 in]{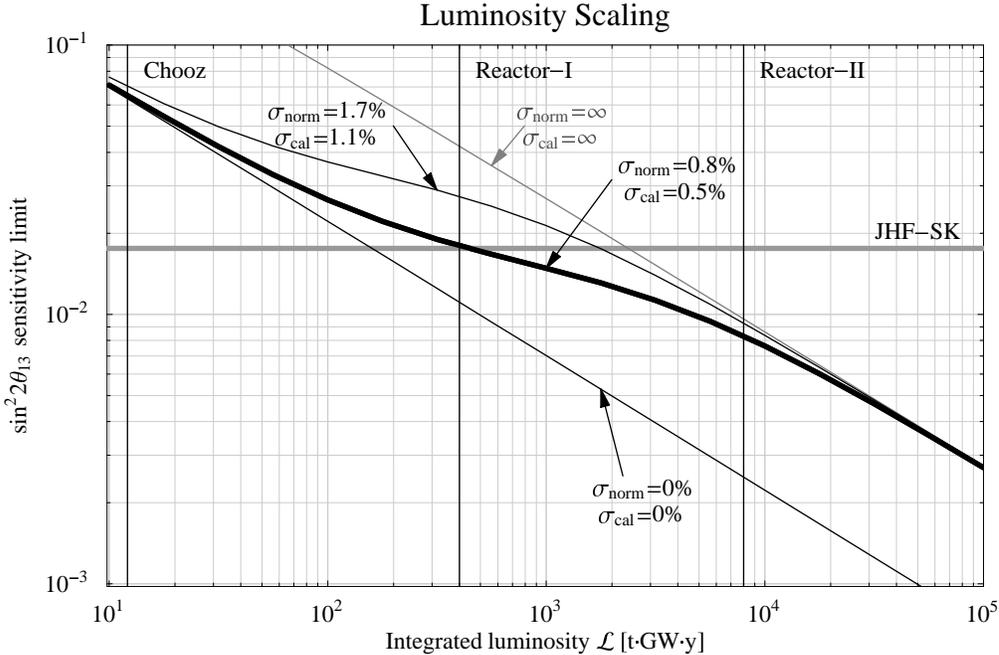}
\vspace{0.15 in}
\caption[]{Projected sensitivity (90\% C.L.) of 
future reactor neutrino experiements to $\theta_{13}$
as a function of integrated luminosity \cite{Huber}. The authors assumed 
$\Delta m_{31}^2 = 3.0 \times 10^{-3}$~eV$^2$ and studied scenarios 
with various systematic errors
for flux normalization ($\sigma_{\rm norm}$) and energy calibration ($\sigma_{\rm cal}$).
The JHF-SK line is the projected sensitivity of the future JPARC-SK experiment 
(see {\it 4.3.2} below),
Reactor-I scenario is 400 GW-ton-years, and Reactor-II scenario is 8000 GW-ton-years.}
\label{fig:lindner}
\end{figure}

Fig.~\ref{fig:lindner} shows the results of 
a general study of the potential of such experiments
\cite{Huber}. With reasonable systematic errors ($< 1\%$) 
for normalization and energy calibration
it is apparent that about 400 GW-ton-years 
of luminosity would provide a determination of 
$\sin^2 2\theta_{13}$ with sensitivity better than 0.02. 
In addition, the figure shows interesting
behavior at higher luminosity ($\sim 8000$ GW-ton-years) 
where the sensitivity improves as 
$1/\sqrt{\mathcal L}$ due to the high statistical precision 
in determination of the relative
spectral shapes at the near and far detectors.
(The curve with $\sigma_{norm}$ and $\sigma_{cal} = \infty$
describes the situation where the absolute efficiency and
energy calibration of the near and far detectors becomes
irrelevant, but the relative efficiencies and energy calibrations
must be still tightly controlled.) 
Although the realization of siting KamLAND-scale detectors 
near large power reactor plants would be very challenging,
the increased sensitivity to $\sin^2 2\theta_{13}$
at such high luminosities would be of great interest.

\subsubsection{Long Baseline Accelerator Experiments}

\begin{figure}[t]
\includegraphics[width=5.0 in]{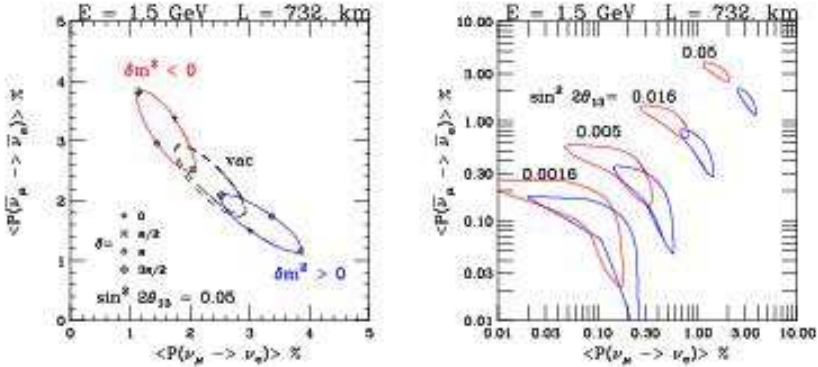}
\caption[]{Oscillation probabilities for neutrino and antineutrino 
oscillations \cite{biprob} for the NUMI experiment parameters \cite{NUMI}. 
The different loops correspond to various values of $\sin^2 \theta_{13}$
and normal and inverted hierarchy. The variation of 
the CP-violating phase $\delta$ traces the loops.}
\label{fig:NUMI}
\end{figure}

While the reactor neutrino studies at $\sim 2$~km 
have the potential to establish a non-vanishing
mixing and a numerical value for $\theta_{13}$, 
further studies related to the neutrino mass
hierarchy and the role of $CP$ violation will 
require mounting new long baseline accelerator
experiments (see Sections {\it 2.4} and {\it 2.5}).
The potential for studying these phenomena is evident in Eq.~(\ref{e:longb}) and
illustrated in Fig.~\ref{fig:NUMI}
where one can see the effects of varying the CP-violation parameters and 
changing the mass hierarchy.
There is a great deal of activity in this subject at the moment, and a variety
of proposed scenarios are under discussion.

The main features of the next generation long baseline accelerator experiments include:
\begin{itemize}
\item[1.)] ability to detect the $\nu_\mu \rightarrow \nu_e$ process as the major goal,
\item[2.)] $L/E_\nu \sim 500$~km/GeV to optimize the oscillation probability for 
$\Delta m_{23}^2 \sim 3 \times 10^{-3}$~eV$^2$,
\item[3.)] propagation of the $\nu$ beam through the earth, resulting in sensitivity
to matter effects,
\item[4.)] possibility to vary $L/E_\nu$ by adjusting the focusing horn, target position,
and/or detector location,
\item[5.)] possibility to switch to $\bar \nu_\mu$.
\end{itemize}
A significant new concept, the ``off-axis'' neutrino beam \cite{BNL_E889}
(combined with a substantial increase in the
neutrino flux and often referred to as a ``super-beam''), appears to be a
very attractive option for these experiments. 
By positioning the detector off the symmetry
axis (at a so-called ``magic'' angle), the neutrino energy 
becomes stationary with respect to the
pion energy and an essentially monoenergetic neutrino beam is obtained. 
Although generally some loss
of flux results from the off-axis geometry, 
the advantages include a tuneable narrow-band beam
with significant suppression of the higher energy tail and 
very low $\nu_e$ contamination. Thus one can selectively scan a range of
$\Delta m^2$ to measure the dependence of the oscillation 
probability with reduced background
rates (both due to the low $\nu_e$ flux and the suppressed $\pi^0$ production from higher
energy neutral current events). This method has been adopted 
for both the JPARC-SK experiment and the Fermilab NUMI proposal.

The JPARC facility \cite{JHF} will be a high luminosity 
(0.77 MW beam power) 50 GeV proton synchrotron
with neutrino production facility aimed about 
$2^\circ$ off-axis from the Super-Kamiokande
detector ($L \simeq 295$~km). The resulting low-energy 
($\sim 0.7$~GeV) neutrino beam at the
Super-K site will enable sensitivity to 
$\sin^2 2\theta_{13} \sim 6 \times 10^{-3}$ after 
about 5 years running. It is 
expected that the experiment would start in 2008, 
and upgrades involving siting a 1Mt detector
and increased beam intensity are envisioned for the future, 
with potential sensitivity to
$\sin^2 2\theta_{13} < 1.5 \times 10^{-3}$ 
and $CP$-violating phase $\delta \sim \pm 20^\circ$.

The ``Off-axis NUMI'' proposal \cite{NUMI} 
is to site a new detector either near Soudan at 
$L \sim 715$~km or at a further site in Canada 
at $L \sim 950$~km, at an angle of about 14 mrad
with respect to the beam axis. The neutrino beam energy 
will be about 2 GeV, and with a 20 kt detector
(still under development) the sensitivity would be 
comparable to the JPARC-SK experiment, but the
higher energy of this neutrino beam improves the sensitivity 
to the sign of $\Delta m_{23}^2$ 
(i.e., normal vs. inverted hierarchy).

There is also discussion of a plan \cite{BNL-NUSL}
to upgrade the AGS at Brookhaven to higher beam power and construct
a neutrino beam aimed at the proposed new underground laboratory (NUSL) 
in Lead, South Dakota or some other underground location at a similar
distance. 
This would afford the opportunity to perform measurements 
with a large ($~0.5$ Mt) detector at a distance
of $L \sim 2500$~km, as well as with a closer off-axis detector 
($L \sim400$~km). Such measurements
could convincingly demonstrate the oscillatory behavior 
of the flavor transformations with high statistics.

\subsection{Future Direct Mass Measurements}

\begin{figure}[t]
\includegraphics[angle=-90,width=5.5 in]{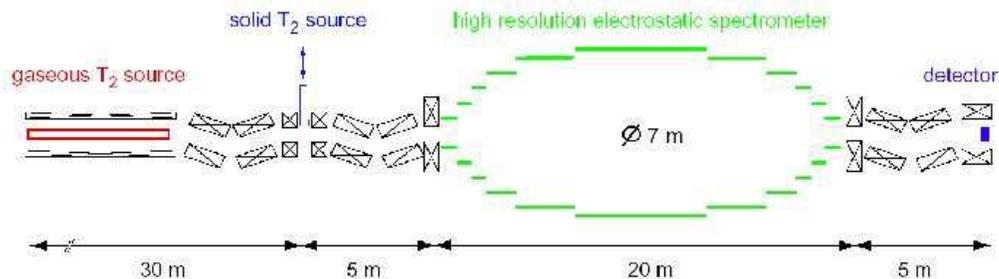}
\caption[]{Conceptual design for the KATRIN experiment for higher-precision
measurements of the neutrino mass using the endpoint 
of the tritium $\beta$-decay spectrum.}
\label{fig:katrin}
\end{figure}

A new experiment to study the tritium $\beta$-endpoint spectrum, 
KATRIN, is under development \cite{katrin}. 
In order to improve the sensitivity to neutrino masses 
by an order of magnitude it is necessary
to both reduce the energy resolution of the spectrometer 
(to $\sim 1$~eV) and increase the tritium source 
strength (to improve statistical precision). The
basic strategy for achieving these goals is to scale up the previous spectrometer
design used in the successful Mainz and Troitsk experiments
to a larger physical size, as shown in Fig.~\ref{fig:katrin}. The larger size enables 
a higher ratio of magnetic fields $B_{\rm max}/B_{\rm min}$ 
to improve the energy resolution
and a larger source acceptance to increase the luminosity. 
The previous spectrometers were 1-1.5~m in diameter 
and the new KATRIN design is 7~m diameter. 
Both a windowless gaseous tritium source and 
a condensed source will be utilized to enable studies
of potential systematic effects associated with 
the different source methods. The goal for the tritium
source is a  column density of $5 \times 10^{17}$ molecules/cm$^2$ 
and a maximum accepted
take-off angle of $\theta_{\rm max} = 51^\circ$. 
These parameters will enable sensitivity to
neutrino masses in the sub-eV range, hopefully down to $\sim 0.35$~eV.

\subsection{Plans for Double Beta Decay}

\begin{figure}[b]
\begin{center}
\includegraphics[width=3. in]{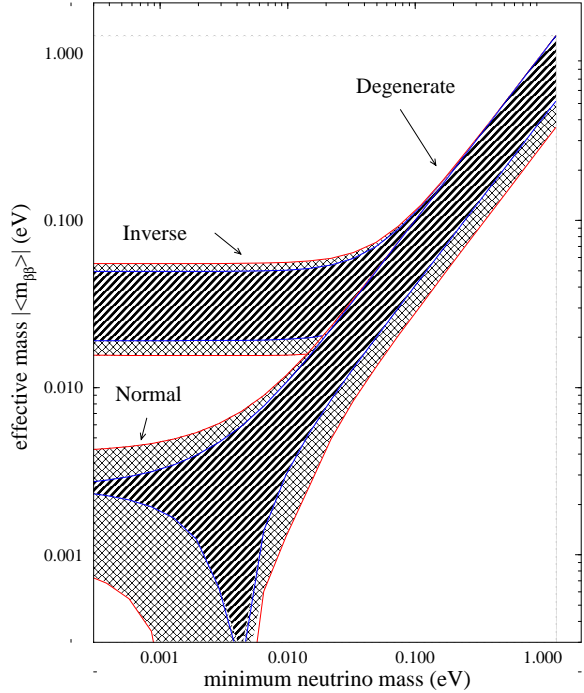}
\medskip
\caption{ Dependence of the effective Majorana mass $\langle m_{\beta\beta} \rangle$
derived from the rate of neutrinoless double beta decay
($1/T_{1/2}^{0\nu} \sim \langle m_{\beta\beta} \rangle^2$) on the absolute
mass of the lightest neutrino. The stripes region indicates the range related
to the unknown Majorana phases, while the cross hatched
region is covered if one $\sigma$ errors on the oscillation parameters
are also included. The arrows indicate the three possible neutrino mass
patterns or ``hierarchies''.
}
\label{fig:bbmass}
\end{center} 
\end{figure}

In contrast to the future direct neutrino mass measurements 
described above in {\it 4.4}, the
next generation of double beta decay searches is 
poised to address the mass scale indicated by
the atmospheric $\nu$ measurements of Super-Kamiokande
($\Delta m_{32}^2 \simeq 2.0 \times 10^{-3}$~eV$^2$) and
perhaps even approach the scale of the solar $\nu$ solution
($\Delta m_{12}^2 \simeq 1\times 10^{-4}$~eV$^2$).
For the case of normal hierarchy ($m_1 \sim m_2 \ll m_3$), 
there is a rather firm prediction for
the effective mass relevant to $\beta \beta$ decay:
\begin{equation}
m_{\beta \beta}^{2} \simeq {\rm Max} \left[
| U_{e3}^2 m_3 |^2, ~|\sin\theta_{12}^2 m_2|^2 \right]
 \sim 10^{-5} {\rm eV}^2  ~.
\end{equation}
This mass scale of less than 3 meV is difficult to 
address in the near future, but if nature
actually chooses the inverted hierarchy 
($m_1 \sim m_2 \gg m_3$), then the $\beta \beta$ prediction
becomes
\begin{equation}
m_{\beta \beta}^{2} \simeq | U_{e1}^2 m_1 |^2 + | U_{e2}^2 m_2 |^2 
\sim  m_1^2   \sim 10^{-3} {\rm eV}^2  ,
\end{equation}
corresponding to an effective mass scale of about 30 meV. 
For the degenerate neutrino mass pattern 
($m_1 \sim m_2 \sim m_3 \gg \sqrt{\Delta m_{32}^2}$)
the effective mass is larger than $\sim$ 50 meV, constrained  from above by the 
mass limit from the tritium beta decay.

The relation between the effective mass in $\beta\beta$ decay and the mass of the
lightest neutrino, evaluated for the solar $\nu$ solution (see Table~1), 
is shown in  Fig. \ref{fig:bbmass}.  

\begin{table}[h]
\caption{Proposed or suggested future $0\nu \beta\beta$
experiments  separated into two groups based on the
magnitude of the proposed isotope mass$^a$.
 }
\label{tab:0nufut}

\renewcommand{\arraystretch}{0.9}
\begin{tabular}{|lclcc|}
\hline \\
       &            &      &  Sensitivity to & range of        \\
Experiment      &       Source     & Detector Description     & 
$T_{1/2}^{0\nu}$ (y) &   $\langle m_{\beta\beta} \rangle$ (meV)   \\ \hline
COBRA\cite{ZUB01}          &$^{130}$Te  &       10 kg CdTe semicond.       & $1 \times 10^{24}$  & 700-2400 \\
DCBA\cite{ISH00}           &$^{150}$Nd  &       20 kg $^{enr}$Nd layers   & $2 \times 10^{25}$ & 35-50   \\
& & between tracking chambers & & \\
NEMO 3\cite{NEMO3}         &$^{100}$Mo  &       10 kg of   $0\nu \beta\beta$ isotopes  & $4 \times 10^{24}$ & 270-1000   \\
& & (7 kg Mo)  with tracking  & & \\ \hline
CAMEO\cite{BEL01}          &$^{116}$Cd  &       1 t  CdWO$_4$ crystals     & $> 10^{26}$  & $>$ (70-220) \\
& & in liq. scint. & & \\
CANDLES\cite{KIS01}        &$^{48}$Ca   &       several tons of CaF$_2$     & $1 \times 10^{26}$ & 160-300     \\
& & crystals in liq. scint. & & \\
CUORE\cite{AVI01}          &$^{130}$Te  &       750 kg TeO$_2$ bolom.    & $2 \times 10^{26}$   & 50-170 \\
EXO\cite{EXO00}            &$^{136}$Xe  &       1 t $^{enr}$Xe TPC      & $8 \times 10^{26}$ & 50-120  \\
GEM\cite{ZDE01}            &$^{76}$Ge   &       1 t $^{enr}$Ge diodes     & $7 \times 10^{27}$  & 15-50   \\
& & in liq. nitrogen & & \\
GENIUS\cite{KLA01b}        &$^{76}$Ge   &       1 t 86\% $^{enr}$Ge diodes  & $1 \times 10^{28}$ & 13-42  \\
& & in liq. nitrogen & & \\
GSO\cite{DAN01,WANGS01}    &$^{160}$Gd  &       2 t Gd$_2$SiO$_5$:Ce crystal  & $2 \times 10^{26}$ & 65  \\
& & scint. in liq. scint. & & \\
Majorana\cite{MAJ01}       &$^{76}$Ge   &       0.5 t 86\% segmented     & $3 \times 10^{27}$ & 24-77 \\
& & $^{enr}$Ge diodes  & & \\
MOON\cite{EJI00}           &$^{100}$Mo  &       34 t $^{nat}$Mo sheets   & $1 \times 10^{27}$ &  17-60 \\
& &  between plastic scint. & & \\
Xe\cite{CAC01}             &$^{136}$Xe  & 1.56 t of $^{enr}$Xe         & $5 \times 10^{26}$  & 60-150  \\
& & in liq. scint. & & \\
XMASS\cite{XMASS}          &$^{136}$Xe  &  10 t of liq. Xe      & $3 \times 10^{26}$  & 80-200 \\ \hline

\end{tabular}

\medskip

$^a$ Adopted from \cite{EV}. The $T_{1/2}^{0\nu}$ sensitivities are those  estimated
by the collaborators but scaled for 5 years of data taking.
These anticipated limits should be used with caution since
they are based on assumptions about backgrounds for experiments
that do not yet exist. Since some proposals are more conservative than others in their
background estimates, one should refrain from using this table to contrast
the experiments. The range of
the effective masses $\langle m_{\beta\beta} \rangle$ 
reflects the range of the calculated
nuclear matrix elements (see Table 2 of Ref. \cite{EV}). Again, caution should be used
since for some nuclei, in particular the deformed $^{150}$Nd and $^{160}$Gd, only
few calculations exist and the approximations are even more severe than in the other
cases.

\end{table}

Present estimates of the nuclear matrix elements \cite{EV}
involved in the $\beta \beta$ process indicate that, 
with of order several tons of enriched
material, experiments could reach this interesting mass range.

There are many proposed
experiments to address this issue in the near future. 
Most are still in the development stage,
and of course the issue of backgrounds is critical in every case.
Since the source mass is $\sim$ 100 times larger than in the 
present experiments, the background per unit mass must be correspondingly
smaller. Different proposals approach this issue differently, and in
all cases substantial R\&D is required. 
Nevertheless, it appears that
several experiments will be mounted during the next decade with 
the goal of studying
$0\nu \beta \beta$ with sensitivity below 30 meV.
The Table 2 lists the proposed experiments we are aware of,
together with the claimed sensitivity to the $0\nu\beta\beta$
halflife, and its crude translation into the sensitivity
to the Majorana mass.

Among the listed experiments, a few relatively small scale searches 
are actually running (COBRA, NEMO3). Others, particularly 
the `ton size', will be mounted in stages with prototypes of 50-200 kg approved and
funded at the present time for several of them 
(one, CUORICINO, the prototype of CUORE, with 40 kg
of natural Te, is already running in Gran Sasso). 
Clearly, the final decision as to which of these many ideas will be realized
depends on the outcome of these prototype installations. Nevertheless,
since they are still $\sim$ 10 times larger than the present
experiments, we can expect substantial improvement in sensitivity
relatively soon.

\subsection{Cosmological Input}

In the near-term future, substantial additional information 
from cosmological observations will become
available that will further constrain the sum of light 
neutrino masses $\sum_f m_{\nu_f}$ compared to the limits
quoted in {\it 3.4}.
Already in progress is the Sloan Digital Sky Survey (SDSS) 
\cite{SDSS} that will map over $10^6$
galaxy redshifts and provide 
relative sensitivity in the relevant region of the power spectrum of $\sim 1 \%$.
When combined with the WMAP CMB data, this should provide sensitivity to 
$\sum_f m_{\nu_f} \sim 0.3$~eV \cite{Hu}. Further in the future, higher precision 
measurements of the CMB anisotropies are expected from the PLANCK mission 
(now expected to be launched in 2007) \cite{PLANCK}. An estimate of the combined 
sensitivity ($1 \sigma$) of the
expected PLANCK data and the SDSS observations to neutrino masses gives the result
$\sum_f m_{\nu_f} \sim 0.06$~eV \cite{Hannestead2}.

\section{Longer-term Outlook}

It is difficult to envision what direction the study
of neutrino mass and oscillations will take in the longer-term.
Thus, we can only offer educated guesses. 

Some of the future research will, obviously, depend on the 
results of the near-term future
described in the preceding section. 
Among the results that might force a revision of the experimental program
are the MiniBooNE tests of the LSND  evidence for the 
$\bar{\nu}_{\mu} \rightarrow \bar{\nu}_e$ oscillations corresponding
to $\delta m^2 > 0.2$ eV$^2$, and the attempts to determine the mixing
angle $\theta_{13}$.

If MiniBooNE confirms the LSND observation, many more experiments will 
be needed because the nature of the neutrino mixing matrix would be
much richer (more than just three active neutrinos or $CPT$ invariance
violation) than envisioned
so far and summarized in Table 1.
Also, if it turns out that the mixing angle $\theta_{13}$ is relatively
large, perhaps close to the present upper limit, the study of the 
possible $CP$ invariance violation in the neutrino sector will become
considerably easier and will proceed faster
than if $\theta_{13}$ is very small.

In addition, neutrino astrophysics will undoubtedly advance. Further studies
of the solar neutrino spectrum will be conducted, and in particular the
low energy part, the flux of the $^7$Be and $pp$ neutrinos, will be
likely 
determined in live 
counting experiments. These measurements will be valuable not
only for the further refinement of the determination of the oscillation
parameters $\Delta m_{12}^2$ and $\theta_{12}$, but also for better
understanding of solar energy production. 

Moreover, with several experiments on-line, 
we can hope that in the near future the next galactic core collapse
supernova will be observed by a variety of neutrino detectors.
If that happens, the observation of SN1987A \cite{1987A} which launched
this field of neutrino astrophysics, will be exceeded many times,
and all components of the supernova neutrino spectrum will be observed.
Again, such observations would not only advance our knowledge of neutrino
physics, but will offer a unique insight in the physics of the stellar
core collapse.  

We also anticipate that during the next decade other applications of
neutrino physics will become reality. For example, the 
observation of `geoneutrinos', i.e. the $\bar{\nu}_e$ emitted by the
decay of radioactive U and Th series in the Earth crust 
\cite{KamLAND,geo1,Fiorentini} would enable determination
of the corresponding radiogenic heat generation and provide
an important contribution to
geophysics. Another application is the measurement of the relic 
supernova flux: the diffuse flux resulting from past supernovae in
the surrounding space, roughly up to redshift $z = 1$. By measuring
this flux one can determine the average supernova rate over a substantial
part of the visible universe (for an estimate
of the rate see e.g. \cite{relic1}). 

At present, neutrino telescopes have not yet seen the TeV and higher
energy  neutrinos which likely accompany the production of cosmic
rays with such energies. This might change in near future
when the relatively small underground detectors are supplemented
by much larger ones in open water or ice. Observation of the direction
of high energy neutrinos would considerably advance the study 
of the origin of high
energy cosmic rays \cite{learned}. It will also make it possible 
to search for
neutrinos from annihilation of the so far hypothetical
dark matter particles (WIMPs). It is beyond the scope of this review
to describe the projects being built or proposed (see \cite{Spiering}
for a recent review). 

Naturally, the whole field of particle physics will advance
at the same time. With the launch of LHC one can imagine that the 
existence of the Higgs boson will be experimentally confirmed, and its
properties will be, at least crudely, determined.
One can also hope that the quest for the understanding of the two
main paradigm of present day physics, the nature of the `dark matter'
and `dark energy', responsible for most of the energy density of the
Universe, will be advanced.

These more general areas of particle physics could be, in fact, intimately
related to the quest for neutrino mass and oscillations.
Taking the see-saw formula ${\mathcal M_L = \mathcal M_D}^2/{\mathcal M_H}$, 
and using for
${\mathcal M_L}$ the neutrino mass scale $\sim$10 meV and for
${\mathcal M_D}$ the electroweak symmetry breaking mass scale $\sim$ 100 GeV,
we arrive at  ${\mathcal M_L} \sim 10^{15}$ GeV, i.e. close to the GUT
unification mass scale. If that relation could be firmed up, we will
arrive at another determination of that important mass scale.
The relation of the neutrino mass scale to the `dark energy' is based
on the, perhaps accidental, numerical coincidence. Remembering that
the energy density of the `dark energy' is about 
$0.7 \Omega_c \sim 3.5\times 10^3$ eV/cm$^3$ and rewriting it in
the `natural units' as $\epsilon^4/(\hbar c)^3$ we arrive at
the dark energy mass scale $\epsilon \sim$ 2 meV, close to the
neutrino mass scale, and unlike any other mass scale in
particle physics.  (For recent
attempts to relate these two mass scales see e.g., \cite{Nelson}.)

Future large scale projects to generate powerful new beams of neutrinos
are already envisioned.
To explore the possibility of $CP$ violation in the lepton sector
one would like to have a well-understood and collimated neutrino
beam of a well defined flavor and energy spectrum. Such a beam 
could be aimed
at a distant large detector. Recently it has been pointed out
that technology exists to construct pure $\nu_e$ and $\bar{\nu}_e$
beams of the required properties. Accelerating radioactive ions,
$^6$He ($E_0 = 3.5$ MeV, $T_{1/2} = 0.8$s, produces $\bar{\nu}_e$)
or $^{18}$Ne ($E_0 = 4.45$ MeV, $T_{1/2} = 0.8$s, produces $\nu_e$)
to $\gamma \simeq 100$, would produce beams of precisely known
energy profile extending to $2 \gamma E_0$, and collimated to
$1/\gamma$. The oscillation signature in the far away detector
would be the appearance of $\mu^+$ or $\mu^-$ \cite{betabeam}.
It is expected that $O(10^{18})$ decays per year can be achieved.

The beams of $\nu_{\mu}$ and $\bar{\nu}_{\mu}$ as well as
$\nu_e$ and $\bar{\nu}_e$ neutrinos
could be obtained in a neutrino factory where accelerated muons
are stored in a ring with long straight sections. Such a beam will
produce neutrino beams with equal mixtures of $\bar{\nu}_{\mu}$
and $\nu_e$ if $\mu^+$ are stored,  and $\nu_{\mu}$ with $\bar{\nu}_e$ 
if $\mu^-$ are stored. It is expected that neutrino factories
could provide $O(10^{20})$ useful muon decays per year. Neutrino factories
would thus provide beams with small systematic uncertainties in the
beam flux and spectrum. $CP$ violation, and determination of 
$\sin^2 2\theta_{13}$
even if it is as small as $10^{-4} - 10^{-5}$ can be achieved with
such beams and baselines of several thousand km \cite{nufact}.

One indication of the importance of the study of neutrinos is 
the recent report ``Facilities for the Future Science, 
A Twenty-Year Outlook'' (see http://www.sc.doe.gov/). It
lists two projects, among the 28 listed in all
fields supported by the US Department of Energy,
relevant to the present topic, the study of double beta
decay in underground detectors treated among the mid-term 
projects and Super Neutrino Beam treated as a far-term project.

Clearly there are high expectations in the community 
of particle physicists that experiments
with neutrinos will continue to provide new insights 
and substantial progress in the coming
decades. Construction and operation 
of the required facilities will be challenging, but the
potential rewards are great, and a large 
and talented community of physicists is poised to
embark on this unique and interesting journey. 

\section{Acknowledgements}

We would like to thank Felix Boehm for encouragement and inspiration 
resulting from his long and pioneering search for neutrino oscillations
at nuclear reactors. We thank John Beacom and Gerry Garvey for reading the manuscript and 
for helpful comments.
This work was supported in part by the US DOE Grant DE-FG03-88ER40397.

\end{document}